\documentclass[aps,pre,twocolumn,superscriptaddress,showpacs,longbibliography]{revtex4-1}
\usepackage{graphicx}
\usepackage{subcaption}
\usepackage{amssymb,amsmath,amsfonts,bm}
\usepackage{hyperref}
\usepackage[dvipsnames]{xcolor}
\hypersetup{urlcolor=blue, colorlinks=true,citecolor=blue,linkcolor=blue} 
\usepackage{float}

\newcommand{\mL}{\mathcal{L}}
\newcommand{\mLv}{\mathcal{\vec{L}}}
\newcommand{\mC}{\mathcal{C}}
\newcommand{\mCv}{\mathcal{\vec{C}}}

\begin{document}
	
\title{Spontaneous layering and power-law order in the three-dimensional fully-packed hard-plate lattice gas.}
\author{Geet Rakala}
\email{geet.rakala@oist.jp}
\affiliation{Okinawa Institute of Science and Technology, 1919-1 Tancha, Onna-son, Kunigami-gun, Okinawa-ken, Japan }
\author{Dipanjan Mandal}
\email{dipanjan.mandal@warwick.ac.uk}
\affiliation{Department of Physics, University of Warwick, Coventry CV4 7AL, United Kingdom}
\author{Soham Biswas}
\affiliation{Departamento de Fisica, Universidad de Guadalajara, Guadalajara, Jalisco, Mexico}
\author{Kedar Damle}
\email{kedar@theory.tifr.res.in}
\affiliation{Department of Theoretical Physics, Tata Institute of Fundamental Research, Mumbai 400005, India}
\author{Deepak Dhar}
\email{deepak@iiserpune.ac.in}
\affiliation{Indian Institute of Science Education and Research, Dr. Homi Bhabha Road, Pashan,  Pune 411008, India}
\author{R. Rajesh}
\email{rrajesh@imsc.res.in}
\affiliation{The Institute of Mathematical Sciences, C.I.T. Campus, Taramani, Chennai 600113, India}
\affiliation{Homi Bhabha National Institute, Training School Complex, Anushakti Nagar, Mumbai 400094,India}

\begin{abstract} 
We obtain the phase diagram of fully-packed hard plates on a cubic lattice.  Each plate covers an elementary plaquette of the cubic lattice and occupies its four vertices, with each vertex of the cubic lattice occupied by exactly one such plate.  We consider the general case with fugacities $s_\mu$ for `$\mu$ plates', whose normal is the $\mu$ direction  ($\mu = x,y,z$). At and close to the isotropic point, we find, consistent with previous work, a phase with long-range sublattice order.  When two of the fugacities $s_{\rm \mu_1}$ and $s_{\mu_2}$ are comparable, and the third fugacity $s_{\mu_{3}}$ is much smaller, we find a spontaneously-layered phase. In this phase, the   system breaks up into disjoint slabs of width two stacked along the $\mu_3$ axis. $\mu_1$ and $\mu_2$ plates are preferentially contained entirely within these slabs, while plates straddling two successive slabs have a lower density. This corresponds to a two-fold symmetry breaking of translation symmetry along the $\mu_3$ axis, leading to ``occupied slabs'' stacked in the layering direction with a separation of one lattice spacing. In the opposite limit, with $\mu_3 \gg \mu_1 \sim \mu_2$, we find a phase with long-range columnar order, corresponding to simultaneous $Z_2$ symmetry breaking of lattice translation symmetry in directions $\mu_1$ and $\mu_2$. The spontaneously-layered phases display critical behaviour, with power-law decay of correlations in the $\mu_1$ and $\mu_2$ directions when the slabs are stacked in the $\mu_3$ direction, and represent examples of `floating phases' discussed earlier in the context of coupled Luttinger liquids and quasi-two-dimensional classical systems. We ascribe this remarkable behaviour to the constrained motion of defects in this phase, and develop a coarse-grained effective field theoretical understanding of the stability of power-law order in this unusual three-dimensional floating phase.
\end{abstract}

\maketitle
\section{Introduction}
\label{Introduction}

Fully-packed dimer models have been studied for several decades. On the one hand, they provide fascinating examples of entropically-driven ordering, closely analogous to Villain's ``order-by-disorder'' phenomena in frustrated magnets~\cite{Villain_orderbydisorder}. On the other hand, on bipartite lattices, they also host highly-correlated liquid phases~\cite{Henley_Coulombphasesreview}. These ``Coulomb phases'' admit a natural description in terms of a field theory for polarization fields. This is a coarse-grained version of a lattice-level description in which each fully-packed dimer configuration of the bipartite lattice is mapped to a divergence-free vector field on links of the lattice~\cite{Youngblood_Axe,Youngblood_Axe_McCoy}. In the two-dimensional cases of square and honeycomb lattices, this effective field theory correctly describes~\cite{Papanikolaou_etal,Alet2006Classical,Patil_Dasgupta_Damle,Desai_Pujari_Damle,Rakala_Damle_Dhar} the power-law columnar ordered state of the fully-packed dimer model on these lattices. In the three-dimensional cases of the cubic and diamond lattice, this correctly predicts that the fully-packed dimer model on these bipartite lattice displays a Coulomb liquid phase with dipolar power-law correlations between the dimers~\cite{Huse_Krauth_Moessner_Sondhi}.

Here we study a fully-packed lattice gas of plates. Each plate covers an elementary plaquette of the cubic lattice and occupies its four vertices. At full-packing, each site of the lattice is occupied by exactly one plate. The key question then is whether such a lattice gas displays correlated liquid phases that could be understood by thinking in terms of tensor-valued analogs of the polarization fields that describe fully-packed dimer configurations. This is a natural speculation since one may view each plate as a pair of parallel dimers on the corresponding plaquette, which translates to two antiparallel dipoles that form a quadrupole with zero net dipole moment. Such a liquid phase would control properties of the corresponding resonating plaquette liquid states of SU(4) magnets~\cite{Xu2008Resonating} in much the same way as the Coulomb liquid phase of the interacting two-dimensional dimer model provides a description of energy correlations of resonating short-range valence bond wavefunctions~\cite{Albuquerque_Alet,Tang_Sandvik_Henley} of SU(N) magnets~\cite{Damle_Dhar_Ramola}.  Motivated perhaps by this natural line of thought, previous work~\cite{Pankov2007Resonating} used Monte-Carlo simulations to study fully-packed hard plates on the cubic lattice, with equal fugacities for plates of all orientations.
It was found however that fully-packed hard plates on the cubic lattice develop long-range sublattice order~\cite{Pankov2007Resonating}, instead of correlated liquid behaviour that could have been described by such a coarse-grained theory of fluctuating quadrupolar tensor fields.

Here, we explore the rich phase diagram of this fully-packed lattice gas of plates in the general case with fugacities $s_\mu$ for ``$\mu$-normal' plates, {\em i.e.} with normal in direction $\mu$ ($\mu = x,y,z$). At and close to the isotropic point ($s_x=s_y=s_z$), we find a phase with long-range sublattice order, {\em i.e.} with two-fold ($Z_2$) symmetry breaking of lattice translation symmetry in all three directions, consistent with the previous results alluded to earlier.  When two of the fugacities $s_{\mu_1}$ and $s_{\mu_2}$ are comparable, and the third fugacity $s_{\mu_{3}}$ is much smaller, we find a spontaneously-layered phase. In this phase, the   system breaks up into disjoint slabs of width two stacked along the $\mu_3$ axis. $\mu_1$ and $\mu_2$ plates are preferentially contained entirely within these slabs, while plates straddling two successive slabs have a lower density. This corresponds to a two-fold symmetry breaking of translation symmetry along the $\mu_3$ axis, leading to ``occupied slabs'' stacked in the layering direction with a separation of one lattice spacing. In the opposite limit, with $\mu_3 \gg \mu_1 \sim \mu_2$, we find a phase with long-range columnar order, corresponding to simultaneous $Z_2$ symmetry breaking of lattice translation symmetry in directions $\mu_1$ and $\mu_2$. The spontaneously-layered phases display critical behaviour, with power-law decay of correlations in the $\mu_1$ and $\mu_2$ directions when the slabs are stacked in the $\mu_3$ direction. The spontaneously-layered phase represents an example of a ``floating phase'', of the type discussed earlier in the context of coupled Luttinger liquids and quasi-two-dimensional classical systems~\cite{Vishwanath_Carpentier,Mukhopadhyay_Kane_Lubensky}. 

To understand the stability of this unusual phase, we develop an effective field-theory that builds on a coarse-grained description used earlier for mixtures of dimers and hard squares on the two-dimensional square lattice~\cite{ramola_columnar_2015}. The basic idea is simple: Consider the limit of a perfectly layered phase, with layering along the (say) $z$ direction. When the layering is perfect, $x$ and $y$ plates are entirely contained within disjoint slabs of width two along the $z$ axis, with no $x$ or $y$ plates straddling two successive slabs. Viewed along the $z$ axis, each occupied slab looks like a system of dimers and hard squares on a two-dimensional square lattice. 

Inter-slab plates that straddle two successive slabs give rise to defects. We argue that these inter-slab plates are bound in pairs in the layered phase.  We develop an effective field theory description in which such pairs of inter-slab plates correspond to quadrupolar couplings between two successive two-dimensional systems. The quadrupolar nature of this couplings renders it irrelevant about the critical fixed point that describes each power-law ordered two-dimensional layer, leading to a stable floating phase.

Thus, this perspective leads us to identify the binding of dipolar defects into quadrupoles as being the key to the stability of this floating layered phase. Microscopically, this arises from the full-packing condition, which leads to constraints on existence and motion of inter-slab defects in this layered phase. The mechanism that stabililizes this phase is thus closely related to the physics of fractonic phases that have attracted a  great deal of attention recently~\cite{Xu2007Bond,Pretko2017Subdimensional,You2020Emergent,You2021FractonicMoessner}. This also suggests that this lattice gas of hard plates is expected to have interesting behaviours in the presence of a small density of vacancies, due to restrictions on the configuration and motion of vacancies in the low density limit. The effect of vacancies has been studied in a parallel work\cite{Dipanjan2021Vacancy} whose results are consistent in part with these expectations.

The rest of this paper is organised as follows: In Sec.\ref{Sec:Model} we first define the model and the order parameters used to identify its phases, and provide an overview of the phase diagram deduced from the results of our Monte-Carlo simulations. Sec.\ref{Sec:Methods} is devoted to a summary of our Monte-Carlo simulation method. In Sec.\ref{Sec:PhasesandTransitions} we present the computational evidence for the phase diagram described in Sec.~\ref{Sec:Model}. Sec.~\ref{Sec:CriticalityandEFT} is devoted to a closer look at the unusual structure of correlations exhibited by the layered phase, and a brief sketch of a coarse-grained field-theoretical description of this phase. Finally, we conclude in Sec.~\ref{sec:Outlook} with a brief discussion of some questions that may be interesting to address in follow-up work.
\begin{figure}
	\includegraphics[width=\columnwidth]{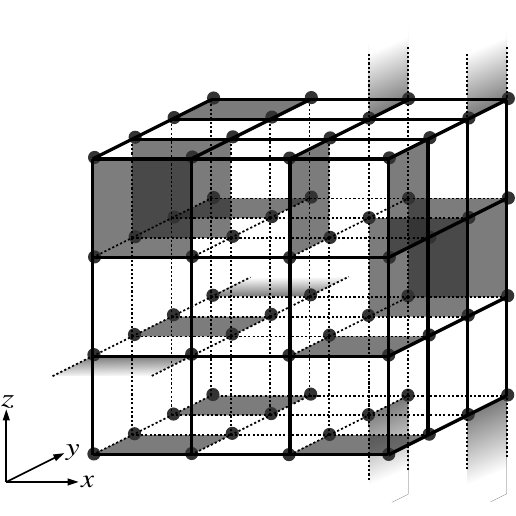}
	\caption{\label{Fig:lattice} 
		A configuration of hard plates at full packing on the $L=4$ cubic lattice with periodic boundary conditions. Each plate occupies exactly four sites of an elementary square plaquette of the cubic lattice, and each site of the cubic lattice is occupied by exactly one such plate in the fully-packed case.
	}
\end{figure}
\begin{table*}
\centering
\begin{tabular}{||c|c|c|c|c|c|c|c||} 
 \hline
 OP & x-columnar & y-columnar & z-columnar&x-layered&y-layered&z-layered&sublattice \\ [0.5ex] 
 \hline\hline
 $\mL_{x}^2$ & 0 & 1 & 1& 1& 0&0&1 \\ 
 \hline
 $\mL_{y}^2$ & 1 & 0&1&0&1&0&1 \\
 \hline
 $\mL_{z}^2$ & 1 & 1&0&0&0&1&1 \\
 \hline
 $\mC_{x}^2$ & 1 & 0&0&0&0&0&1 \\
 \hline
 $\mC_{y}^2$ & 0 & 1&0&0&0&0&1\\
 \hline
 $\mC_{z}^2$ & 0 & 0&1&0&0&0&1 \\
 \hline
 $\omega^2$& 0 & 0&0&0&0&0&1\\ [1ex]
 \hline
\end{tabular}
\caption{Table showing various order parameters and the ordering patterns probed by them. $1$ denotes a nonzero thermodynamic limit for a particular order parameter in that phase, while $0$ indicates that it vanishes in the thermodynamic limit. The order parameters constructed using the alternate double-plate definitions behave in an entirely analogous manner.}
\label{Tab:op}
\end{table*}
\begin{figure}[t]
	\includegraphics[width=\columnwidth]{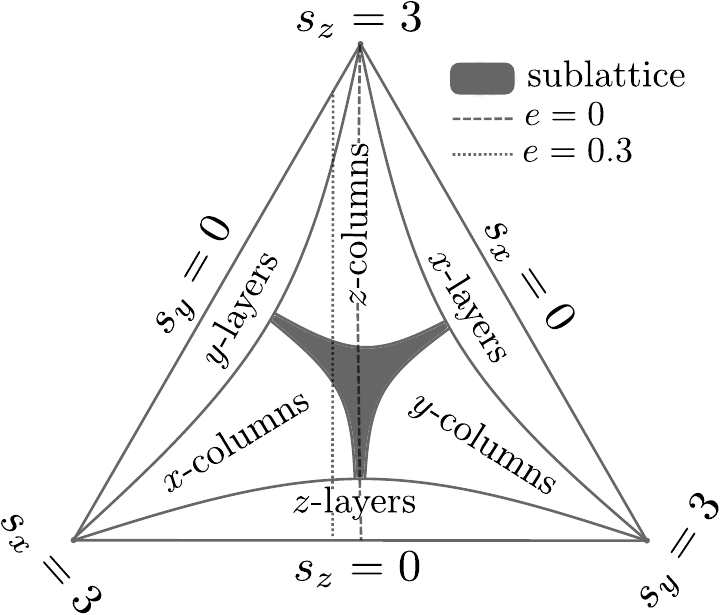}
	\caption{\label{Fig:phase_diagram} 
		A schematic phase diagram inside the fugacity triangle that represents the parameter space of the fully-packed hard square lattice gas. Note that the fugacities $s_\mu$ ($\mu = x,y,z$) satisfy $s_x+s_y+s_z=3$ in our convention. In this convention, the corners of the triangle correspond to the points $s_x=3,s_y=3$ and $s_z=3$ respectively, while the sides correspond to the lines $s_x=0$, $s_y=0$, and $s_z=0$ respectively. The two cuts shown, corresponding to $e=0$ and $e=0.3$ (where $e=s_x-s_y$), are the ones along which our most detailed results have been obtained. There are three kinds of phases: A sublattice-ordered phase, with lattice translation symmetry broken in all three Cartesian directions, layered phases with lattice translation symmetry broken only along one cartesian direction, and columnar ordered phases with lattice translation symmetry broken along two of the three cartesian directions.  
	}
\end{figure}

\section{Model, phase diagram and order parameters.}
\label{Sec:Model}
As already outlined in  Sec.~\ref{Introduction}, we study a fully-packed lattice gas of ``hard'' plates that each occupy all four sites of an elementary plaquette of the cubic lattice, with the proviso that each site of the cubic lattice is occupied by exactly one plate. We consider the general case with fugacities $s_\mu$ for `$\mu$ plates', {\em i.e.} with normal in direction $\mu$ ($\mu = x,y,z$). With periodic boundary conditions, the $L^3$ sites of a $L \times L \times L$ cubic lattice are covered by $L^3/4$ hard plates in every such fully-packed configuration. We assign fugacities $s_{x}$, $s_{y}$ and $s_{z}$ to $x$, $y$, and $z$ plates respectively, with all $s_\mu \geq 0$. We adopt the convention that these fugacities obey the constraint:
\begin{equation}
s_x + s_y + s_z =3.
\end{equation}
The partition function of the system is now given as
\begin{equation}\label{eq:partition_function}
Z = \sum_C s_{x}^{n_{x}} s_{y}^{n_{y}} s_{z}^{n_{z}},
\end{equation}
where the sum is over all the fully-packed configurations of such a cubic lattice. An illustrative example of a configuration of the $L=4$ cubic lattice with periodic boundary conditions is shown in Fig.~\ref{Fig:lattice}.
\begin{figure*}
	\includegraphics[width=2\columnwidth]{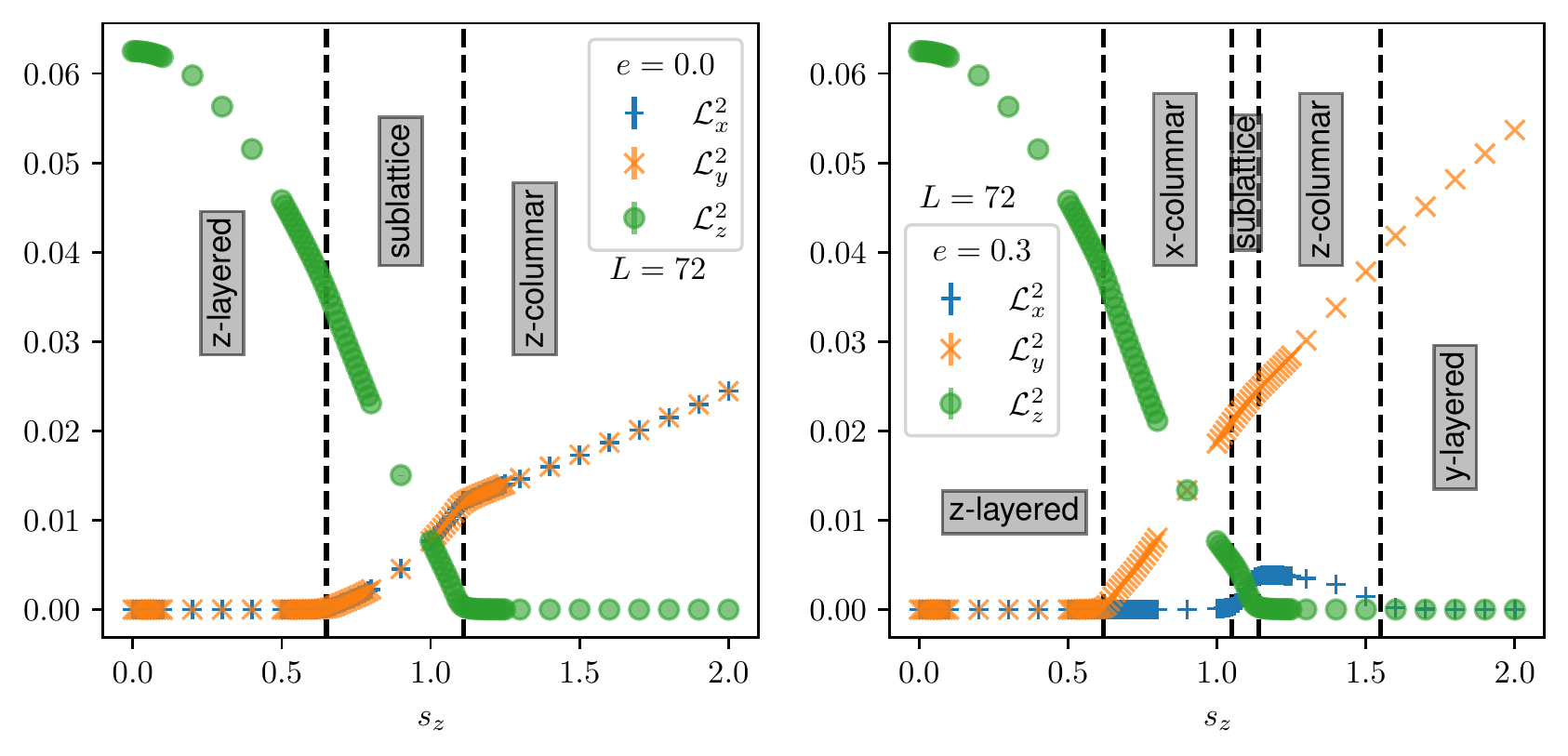}
	\caption{ The behaviour of the three different components of the layering vector $\vec{{\mathcal L}}$ along the two cuts shown in Fig.~\ref{Fig:phase_diagram} provides a convenient way to recognize at a glance the presence of sublattice ordered, columnar ordered, and layered phases in various parts of the phase diagram. The sublattice ordered phase is distinguished by nonzero mean square values for all three components of the layering vector. In a columnar ordered phase, two components of the layering vector have nonzero mean square values, while the mean square value of the third vanishes in the thermodynamic limit. In a layered phase, only one component of the layering vector has a nonzero mean square value.
	}
	\label{Fig:Summaryofscans} 
\end{figure*}

In our computational work, we use three different order parameters to quantitatively probe the nature of the various phases. These are the layering order parameter vector $\mLv$, the columnar order parameter vector $\mCv$ and the scalar sublattice order parameter $\omega$. To define these, it is convenient to adopt a convention whereby each plate is assigned to a unique lattice site $i$ as follows: Any plate occupies four lattice sites. Except on ``boundary plaquettes'' created by the periodic boundary conditions, each plate is assigned to the site $i$ (chosen from these four sites) that has the minimum value of all three cartesian coordinates $x(i)$, $y(i)$ and $z(i)$. On the wrapping plaquettes, this rule is modified in the obvious way for the cartesian coordinate(s) along which periodicity is being imposed. In the fully-packed limit, this implies that $1/4^\textrm{th}$ of the lattice sites are assigned plates and the rest are occupied by plates assigned to an adjacent lattice site.

Using this convention, we define two kinds of occupation variables on each site of the lattice: the single plate occupation $\eta_{\mu}(i)$ takes a value $1$ if a $\mu$ plate is assigned to the site $i$ and 0 otherwise, and the double plate occupation number defined as $\eta_{\mu \mu }(i) = \eta_\mu (i) \eta_\mu( i + \hat{\mu})$, where $\hat{\mu}$ is the unit translation along the positive $\mu$ axis (the definition of $\hat{\mu}$ also incorporates periodic boundary conditions in all three directions).

The layering vectors $\mL_\mu$ and $\mL_{\mu \mu}$ are defined to measure translation symmetry breaking along direction $\mu$. For $\mL_{\mu}$, we have
\begin{eqnarray}
\mL_{x} &=& \frac{1}{L^3}\sum_{i}  l_{x}(i), \nonumber \\
l_x(i) &=& (-1)^{x(i)}(\eta_{x}(i) + \eta_{y}(i) + \eta_{z}(i)),
\label{eq:layering}
\end{eqnarray}
and similarly for $\mL_y$ and $\mL_z$. The layering vector $\mL_{\mu \mu}$ is defined in an entirely analogous manner, with all $\eta_{\gamma}$ in the above replaced by $\eta_{\gamma \gamma}$. 
The corresponding definition of the columnar vectors $\mC_{\mu}$ and $\mC_{\mu \mu}$ is chosen to ensure that these vector order parameters are sensitive to translation symmetry breaking in the two cartesian directions perpendicular to $\mu$. For $\mC_{\mu}$ we have:
\begin{eqnarray}
\mC_{x} &=& \frac{1}{L^3}\sum_{i} c_{x}(i), \nonumber \\ 
	c_{x} &=& (-1)^{y(i)+z(i)} (\eta_{x}(i) + \eta_{y}(i) + \eta_{z}(i)) \; ,
\end{eqnarray} 
and cyclically for the other components $\mC_y$ and $\mC_z$. The layering vector $\mC_{\mu \mu}$ is again defined as above, but with all $\eta_{\gamma}$ replaced by $\eta_{\gamma \gamma}$.

In addition, we define the scalar sublattice order parameter $\omega$ to measure the simultaneous breaking of translation invariance along all three directions:
\begin{equation}
\omega = 27(\mL_x\mL_y\mL_z) \; ,
\label{usefulsublatticedefn}
\end{equation}
where the factor of $27$ is simply a convenient convention. Note that this quantity, unlike the layering and columnar vectors, is not being defined as a sum over a scalar density defined locally on the lattice. Instead, it is simply the product of three components of the layering vector $\mL_{\mu}$. This product is sensitive to the simultaneous breaking of lattice translation symmetry in all three directions of the cubic lattice, and hence probes the presence of sublattice order.
For a convenient summary of the various order parameters and the ordering patterns they probe, see Table.~\ref{Tab:op}.

For each order parameter ${\mathcal O}$, we monitor the $L$ dependence of $\langle {\mathcal O}^2 \rangle$ to determine whether the phase in question is characterized by the corresponding long range ordering behaviour. Since our definitions of all the order parameters include a normalization by a factor of $L^3$ to render it intensive, $\langle {\mathcal O}^2\rangle$ is expected to vanish as $L^{-3}$ in the absence of long range order, tend to a nonzero limit in the presence of long range order, and vanish as $L^{-(d-2+\eta)}$ with $d = 3$ if it has an order parameter correlation function falling off as $1/r^{d-2+\eta}$ with $\eta<2$. Note however that $\langle {\mathcal O}^2 \rangle$ cannot distinguish a power-law ordered phase with $\eta>2$ from a disordered phase, since $\langle {\mathcal O}^2 \rangle \sim 1/L^3$ in both cases, but the two can still be distinguished by measuring the corresponding correlation function of the local order parameter density, except in the case of the sublattice order parameter $\omega$ which is not the sum of a local order parameter density (for further discussion of this point, see Sec.~\ref{sec:Outlook}). Finally, it is also important to realize that $\langle \omega^2 \rangle$ is expected to have a somewhat different fall-off in a phase without sublattice order: If the phase is a columnar ordered phase, one expects a $1/L^3$ fall off, but a phase which is disordered in two cartesian directions and breaks lattice translation symmetry in only one direction is expected to display a $1/L^6$ fall off of $\langle \omega^2 \rangle$. As we will see below, the layered phase that arises in our hard plate system has power-law order in the transverse directions. As a result, we expect $\langle \omega^2 \rangle$ to display a power-law fall off with exponent somewhat smaller than $6$.

For each of these order parameters, we also monitor the Binder ratio $U_\mathcal{O}$:
\begin{equation}
U_{\mathcal{O}} = \frac{\langle \mathcal{O}^4 \rangle}{\langle \mathcal{O}^2 \rangle^2}.
\end{equation}
Clearly, this ratio tends to unity in the thermodynamic limit of an ordered state. In a disordered state, it tends to a value larger than one, which depends on the number of independently fluctuating components that make up ${\mathcal O}$. Thus, the Binder ratio of a scalar order parameter tends to a limiting value of $3$, while that of a two-component vector tends to $2$, and so on.
\begin{figure}[t]
	\includegraphics[width=0.7\columnwidth]{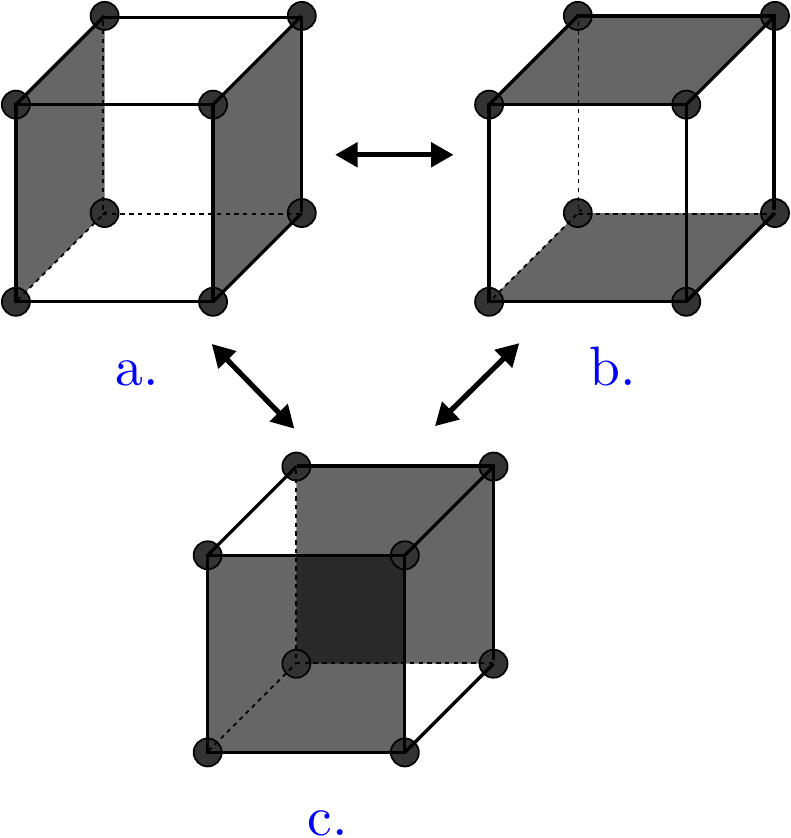}
	\caption{\label{Fig:ppc} 
		Figure showing all the local moves in a ring exchange. 
	}
	\end{figure}

The parameter space of this fully-packed lattice gas is conveniently represented by an equilateral {\it fugacity triangle}, with the three vertices corresponding to $s_x=3,s_y=3$ and $s_z=3$ respectively as shown in Fig.~\ref{Fig:phase_diagram}. The sides of this triangle then correspond to the $s_x=0, s_y=0$ and $s_z=0$ lines. The centroid of this triangle represents the isotropic point where $s_x=s_y=s_z=1$. In our computational work, we have focused on obtaining results along the two cuts shown in Fig.~\ref{Fig:phase_diagram}. A convenient way to summarize our findings along these two cuts is to plot the behaviour of the three components of the layering vector $\vec{{\mathcal L}}$ along these cuts. This is shown in Fig.~\ref{Fig:Summaryofscans}. Supplementing this with results for the behavior of $\langle \omega^2 \rangle$ (discussed separately in Sec.~\ref{Sec:PhasesandTransitions}) and expectations based on the three-fold symmetry of the phase diagram allows us to deduce the structure of the phase diagram in the whole equilateral triangle.  A schematic of the phase diagram thus obtained is displayed in Fig.~\ref{Fig:phase_diagram}.

As is clear from this phase diagram, we find a phase with long-range sublattice order at and close to the isotropic point, {\em i.e.} when all three fugacities are comparable to each other, and the corresponding densities are likewise comparable.  This finding is consistent with a previous study of isotropic fully-packed plates, which found in favour of a sublattice-ordered state~\cite{Pankov2007Resonating}. Our study allows us to go beyond this and study the competition between this sublattice-ordered phase and the layered and columnar-ordered phases for general values of the fugacities. 

Indeed, when two of the fugacities $s_{\rm \mu_1}$ and $s_{\mu_2}$ are comparable, and the third fugacity $s_{\mu_{3}}$ is much smaller, we find a spontaneously-layered phase with $Z_2$ symmetry breaking of lattice-translation symmetry in the $\mu_{3}^{\rm th}$ direction. In this phase, $\mu_1$-normal and $\mu_2$-normal plates preferentially occupy a stack of slabs of width two separated from each other along the $\mu_3$ axis by one lattice spacing. 
In the opposite limit, with $\mu_3 \gg \mu_1 \sim \mu_2$, we find a phase with long-range columnar order, corresponding to simultaneous $Z_2$ symmetry breaking of lattice translation symmetry in directions $\mu_1$ and $\mu_2$. 

To probe the nature of the layered phase in more detail, we study the intra-slab and inter-slab correlation functions of the transverse components of the layering order parameter. To this end, we first define
\begin{eqnarray}
l_{x \perp}(z) &=&\frac{1}{L^2}\sum_{x,y}l_x(x,y,z), \nonumber \\
l_{y \perp}(z) &=& \frac{1}{L^2}\sum_{x,y}l_y(x,y,z).
\end{eqnarray}
In terms of $l_{x \perp}$ and $l_{y \perp}$, we can now define
\begin{eqnarray}
 G(L,\Delta z) &=& \frac{1}{L}\sum_z \langle (l_{x \perp}(z)l_{x \perp}(z+\Delta z) + l_{y \perp}(z)l_{y \perp}(z+\Delta z) \rangle.  \nonumber \\
 &&
 \end{eqnarray}
 In what follows, we will denote $G(z,\Delta z = 0)$ by simply $G(L)$, while keeping the second argument when it is nonzero. In principle, the sum over $z$ in the definition of $G(L,\Delta z)$ should be taken only over ``occupied'' slabs, with the convention that all $z$ plates are assigned to the occupied slabs. However, we have checked that summing over all $z$, which is computationally simpler, gives qualitatively similar results, motivating the definition above.
As an independent check on this way of looking at the intra-slab correlations, we have also measured the $r$ dependence of $g(r)$, defined as the connected intra-slab correlation function of $l_{x}$ and $l_{y}$ within a single slab. Power-law columnar order, with $g(r) \sim 1/r^{\eta}$ within a slab, should correspond to a measured $G(L) \equiv G(L,\Delta z = 0) \sim 1/L^\eta$ if $\eta <2$, and a measured $G(L) \sim 1/L^2$ if $\eta>2$; our results are consistent with this expectation.
\begin{figure}[t]
	\includegraphics[width=\columnwidth]{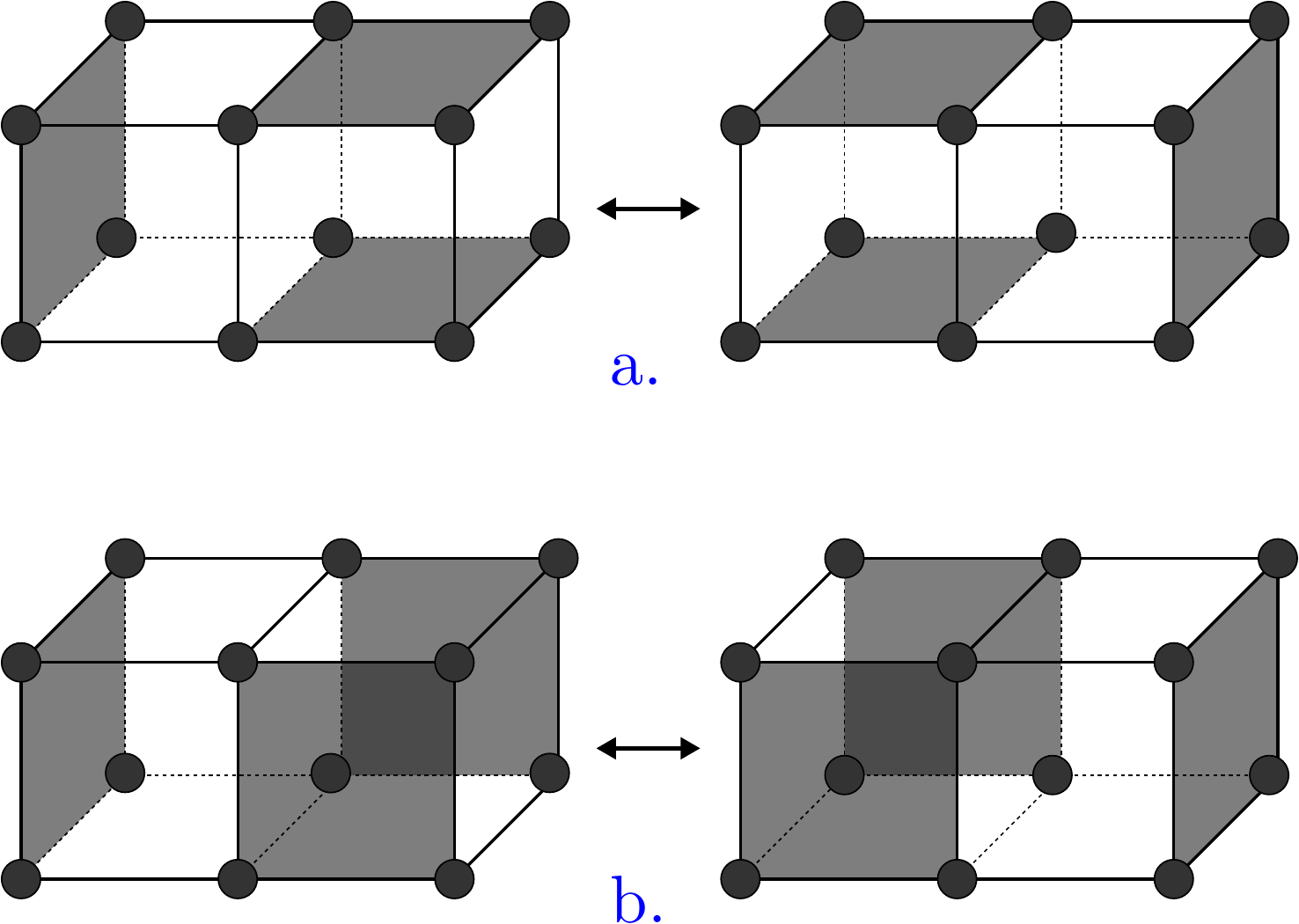}
	\caption{\label{Fig:pshift} 
		Figure showing the local moves in a shift exchange. 
	}
\end{figure}

\section{Computational methods}
\label{Sec:Methods}
Our Monte Carlo simulations use a combination of local updates supplemented by a cluster algorithm to improve the equilibration and reduce autocorrelation times. 
We use two types of local updates: a {\it ring} exchange move, and a {\it shift} exchange move:

{\it Ring exchange:} The ring exchange move, shown in Figure~\ref{Fig:ppc},  relies on the fact that
all eight vertices of an elementary cube can be occupied by two plates that cover two opposite faces of the cube. There are three such ``perfect covers'' of any elementary cube. The idea is to try and exchange one such perfect cover of the cube for another perfect cover chosen with probabilities that satisfy detailed balance. To implement this, one randomly picks an elementary cube of the cubic lattice, and checks if it is of this ``perfectly covered'' or ``flippable'' type. If the answer is yes, then one looks up a previously constructed probability table listing the relative weights in the partition sum of the three different perfect covers of this cube. Using this probability table, one chooses transition probabilities with detailed balance, and replaces the original perfect cover with a fresh one. If a ``no-bounce'' solution of the detailed balance constraints is possible~\cite{Syljuasen_Sandvik}, then the final state is always one of the other two perfect covers. If the weight of one of the three perfect covers is larger than the sum of the other two, then a ``no-bounce'' solution is not possible~\cite{Syljuasen_Sandvik}, and there is a nonzero probability that the final state will be the same as the original perfect cover.

{\it Shift exchange:}  The {\it shift} exchange starts by picking a random pair of adjacent elementary cubes of the lattice sharing a face with each other. Let the shared face have a normal in the $\mu$ direction. If one of the elementary cubes is perfectly covered by a pair of $\nu$ plates, where $\nu \neq \mu$, and the other elementary cube has its remaining four sites occupied by a $\mu$ plate, then the configurations of the two elementary cubes are interchanged (see Fig.~\ref{Fig:pshift}). Since this does not change the orientation of the three plates involved, no probability table is needed, and this shift exchange can be carried out whenever the randomly chosen pair of adjacent elementary cubes host a configuration of this type.

To supplement these two updates and improve autocorrelation times of the Monte Carlo algorithm, particularly in the presence of periodic boundary conditions, one can supplement these basic local updates with two kinds of non-local updates. The first uses a transfer matrix method to update an entire one dimensional tube, keeping the rest of the configuration fixed. This generalizes the approach used previously for simulations of a mixture of hard squares and dimers in two dimensions~\cite{ramola_columnar_2015}. The other is a cluster algorithm that generalizes the pocket cluster algorithm~\cite{krauth_pocket_2003} used previously to study fully-packed dimer models.  Here, we show results obtained using the second approach, {\em i.e.} using a generalization of the pocket cluster algorithm~\cite{krauth_pocket_2003}. In parallel work~\cite{Dipanjan2021Vacancy} that studies the effects of vacancies on our fully-packed lattice gas, we use the alternate transfer matrix scheme~\cite{Dipanjan2021Vacancy}. 

In the cluster update scheme, we rely on the existence of ${\mathcal O}(L)$ different reflection symmetries of the system, each of which square to the identity operation. For this to be the case, periodic boundary conditions are essential. To begin, a reflection is specified by picking a randomly chosen reflection plane. Six kinds of reflection planes, each perpendicular to one of the three cartesian directions and either containing sites of the lattice or bisecting bonds of the lattice, always define valid symmetry operations for general values of fugacities. Of these, we choose to use the $3L$ planes that contain sites of the lattice. In addition, one can consider reflections about ``diagonal'' reflection planes, specified by a normal that lies either in the $xy$ plane and makes an angle of $\pm \pi/4$ with the $x$ and $y$ axis, or does the same in the $yz$ or $zx$ plane. These $6$ kinds of additional reflections are all symmetries of the system only at the isotropic point when all the fugacities are equal. Elsewhere in the fugacity triangle, at most two of them are symmetries. 

The cluster update begins by randomly choosing (with equal probability) one of the possible reflection symmetries of the system, say a reflection about the symmetry plane $\mathcal{S}$. In addition, a plate is chosen at random as the seed. Apart from the physical system, one constructs a ``pocket lattice'', which is empty to begin with. The seed plate is acted on by the reflection operation corresponding to the randomly chosen symmetry plane $\mathcal{S}$. This can change its location, and, possibly its orientation too. Thus transformed, the plate is moved to the pocket lattice and placed in its possibly new location (and possibly new orientation). Next, this transformed plate is moved back to the physical lattice so that it now occupies a transformed location with a transformed orientation in the physical lattice. As a result, it touches some other plates in the physical lattice. These other plates are now removed from the physical lattice, transformed according to the symmetry operation ${\mathcal S}$, and moved to the pocket lattice. 

The rest of the update consists of repeating this process until the pocket lattice is empty. In other words, at each step, we take a plate from the pocket lattice, move it back to the physical lattice so that it occupies a transformed location with a transformed orientation, and then remove the plates it touches in the physical lattice, transform them, and move them to the pocket lattice. Once the pocket lattice is emptied at the end of this process, the physical lattice again has a fully-packed configuration that does not violate any constraints. This new configuration can be accepted with probability one since the weight of the configuration has not changed in this process.
  
We simulate the system using a combination of these local and cluster updates. Each Monte Carlo step (MCS) involves $L^3$ ring exchanges, $L^3$ shift exchanges in the $\hat{x}$, $\hat{y}$ and $\hat{z}$ directions each, and a number of cluster updates, each involving a randomly chosen symmetry plane and a random seed plate (this number is chosen to ensure that a total of ${\mathcal O}(L^3)$ plates are involved in these cluster updates as a whole). The Monte Carlo routine was tested against exact enumeration on a $4 \times 4 \times 2$ periodic lattice with fully-packed hard plates using Martin's backtracking algorithm~\cite{domb1972phase}. 

\begin{figure}
  \begin{subfigure}[b]{\columnwidth}
    \includegraphics[width=\linewidth]{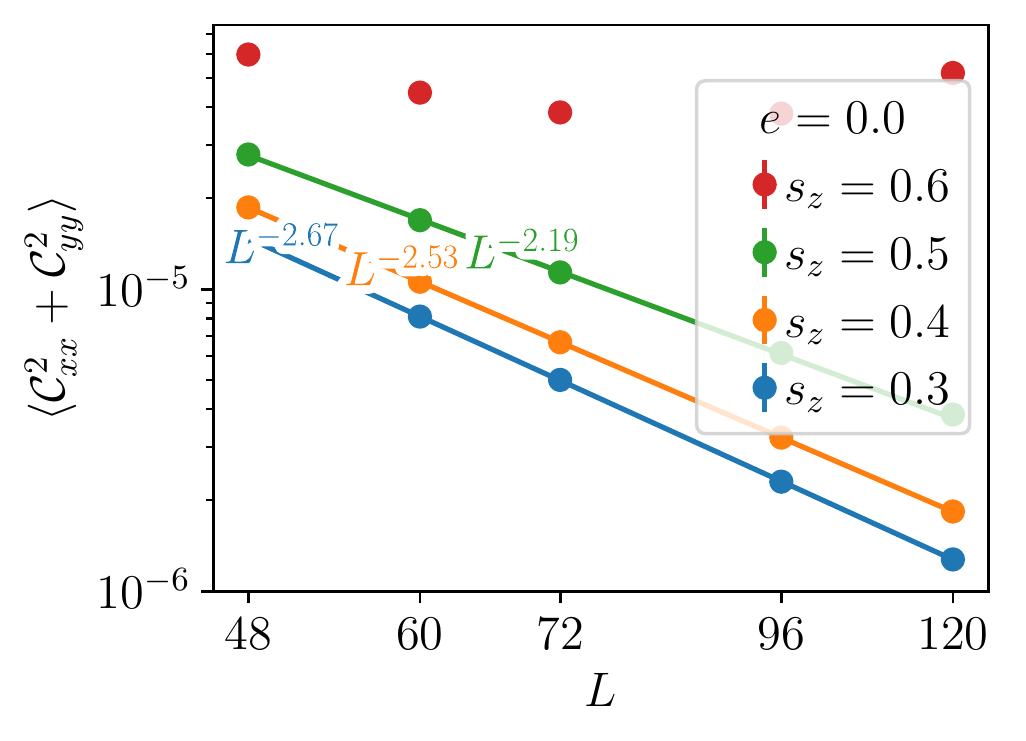}
  \end{subfigure}
  \hfill 
  \begin{subfigure}[b]{\columnwidth}
    \includegraphics[width=\linewidth]{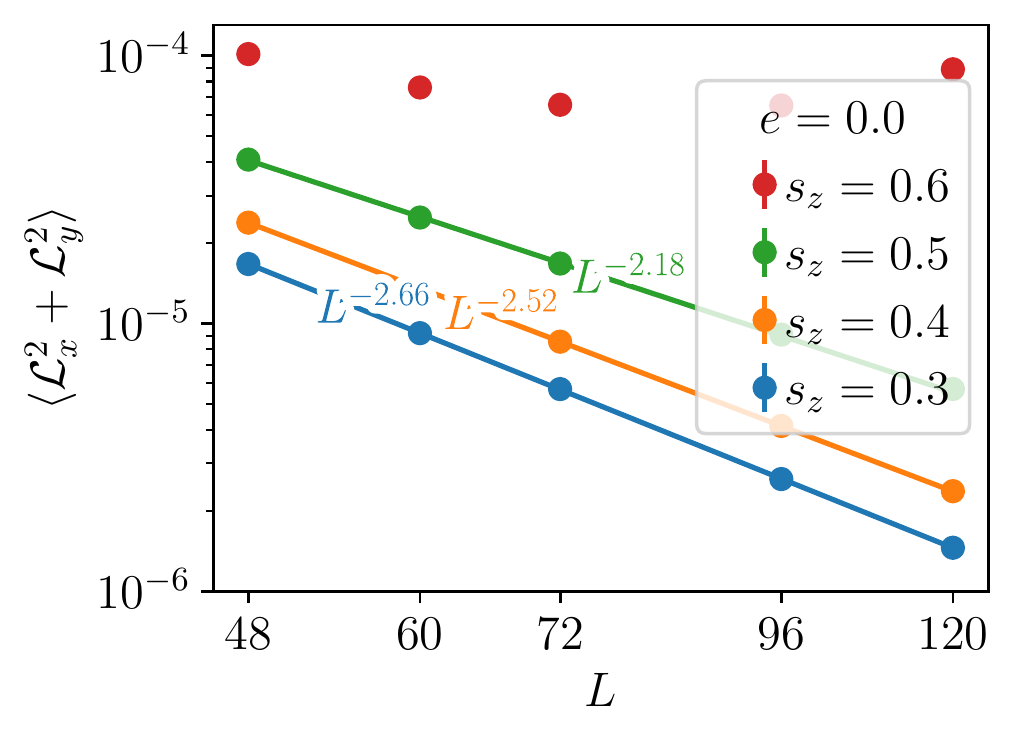}
  \end{subfigure}
  \caption{$\langle C_{xx}^2+C_{xx}^2 \rangle$ (top) and $\langle L_{x}^2+L_{y}^2 \rangle$ (bottom) as a function of $L$ on the $e=0$ cut across the $z$ layered to sublattice order transition for $s_z \in [0.3,0.6]$.}
     \label{Fig:LCperp_e0LS_op}
\end{figure}
 
At each set of fugacity parameters, simulations were carried out on lattices of size $L=48,60,72,96$ and $120$. Simulations at each set of parameter values involved at least 10 random initial seeds, a warm-up of $2 \times 10^5$ MCS for each seed followed by a run length of $2 \times 10^6$ MCS for each seed, with measurements being taken after alternate MCS. $10^3$ such measurements were binned to create one notionally independent measurement. Error bars were estimated by re-binning these notionally independent measurements in sets of $5$, $50$ and $100$ and using the re-binned estimators to calculate three different estimates of the statistical error. The actual error bar assigned to each data point was taken to be the maximum of these three estimates. Independent simulations at each set of parameter values were parallelised using GNU parallel.

\begin{figure}
  \begin{subfigure}[b]{\columnwidth}
    \includegraphics[width=\linewidth]{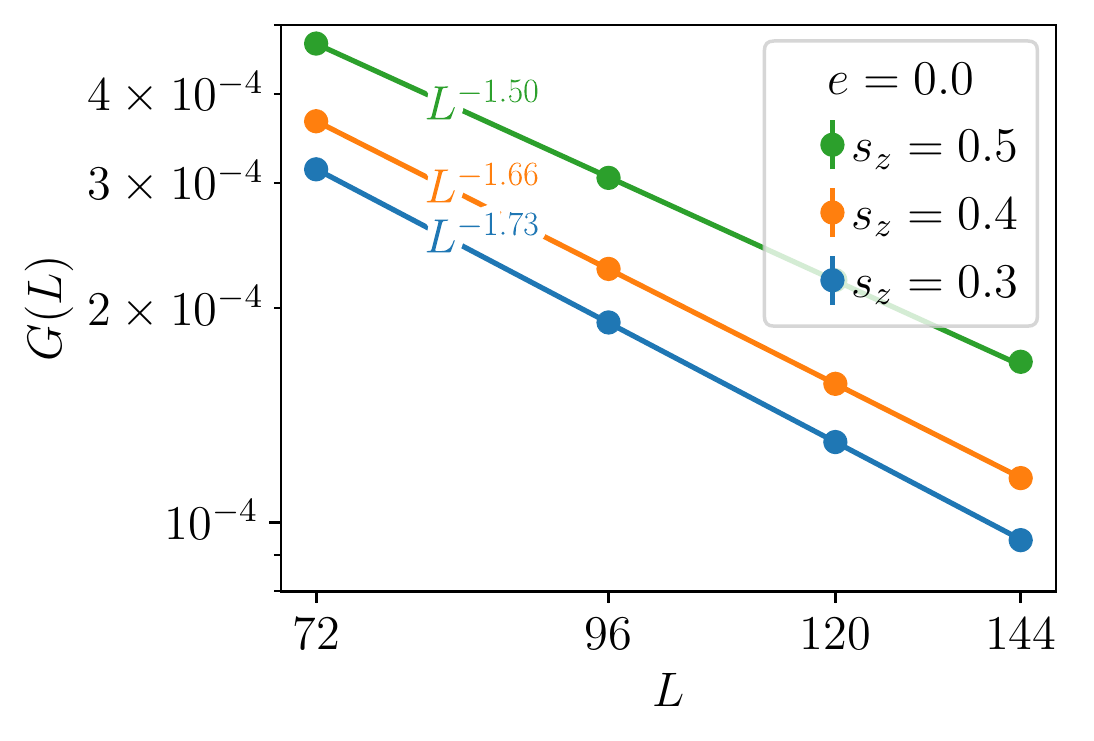}
  \end{subfigure}
  \hfill 
  \begin{subfigure}[b]{\columnwidth}
    \includegraphics[width=\linewidth]{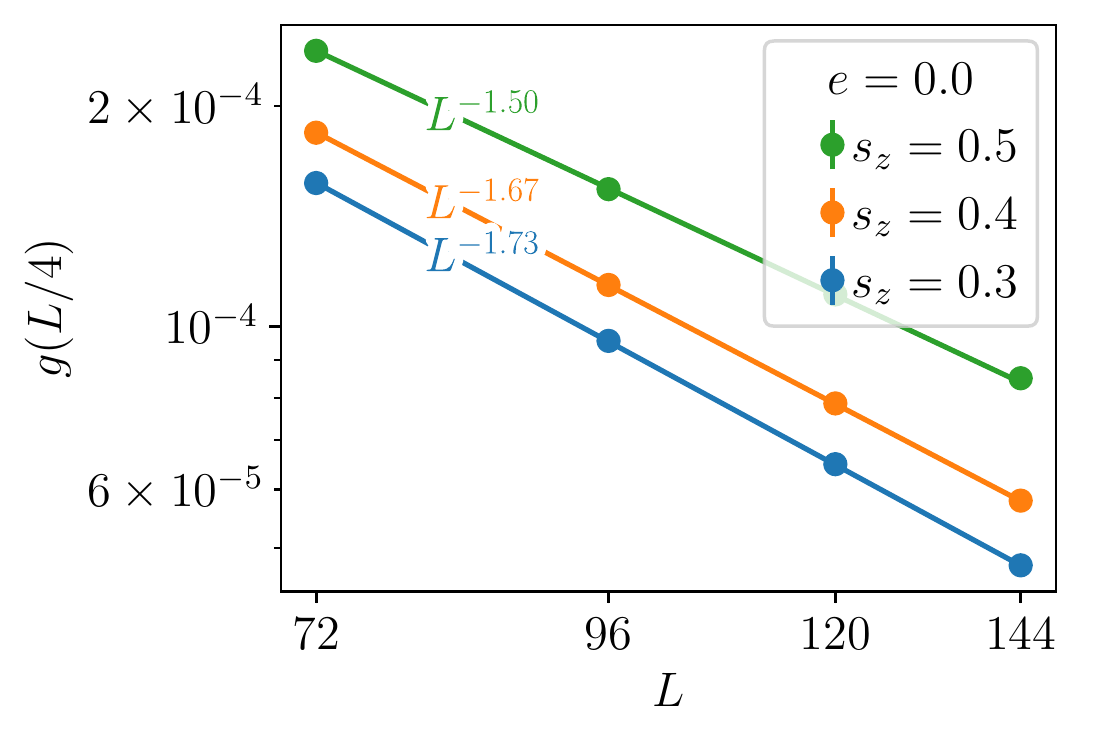}
  \end{subfigure}
  \caption{$G(L)$ (top) and $g(L/4)$ (bottom) as a function of $L$ on the $e=0$ cut for $s_z \in [0.3,0.5]$.}
     \label{Fig:G(L)andg(r)}
\end{figure}

\section{Computational results}
\label{Sec:PhasesandTransitions}
We now turn to a more detailed discussion of our results along the two cuts displayed in Fig.~\ref{Fig:phase_diagram}. With the parameter $e$ defined as $e \equiv s_x - s_y$, these two cuts through the phase diagram correspond to $e=0.0$ and $e=0.3$.
The overview presented in Fig.~\ref{Fig:Summaryofscans} shows the fugacity dependence of the different components of the layering order parameter at a fixed size $L=72$. To pinpoint the transition from one phase to another, and study in more detail the phases themselves, it is of course necessary to study the $L$ dependence of $\langle {\mathcal L}_\mu^2\rangle$ as well as the Binder ratios corresponding to these order parameters. Here, we present some representative results of such a study.

\begin{figure}
\includegraphics[width=\linewidth]{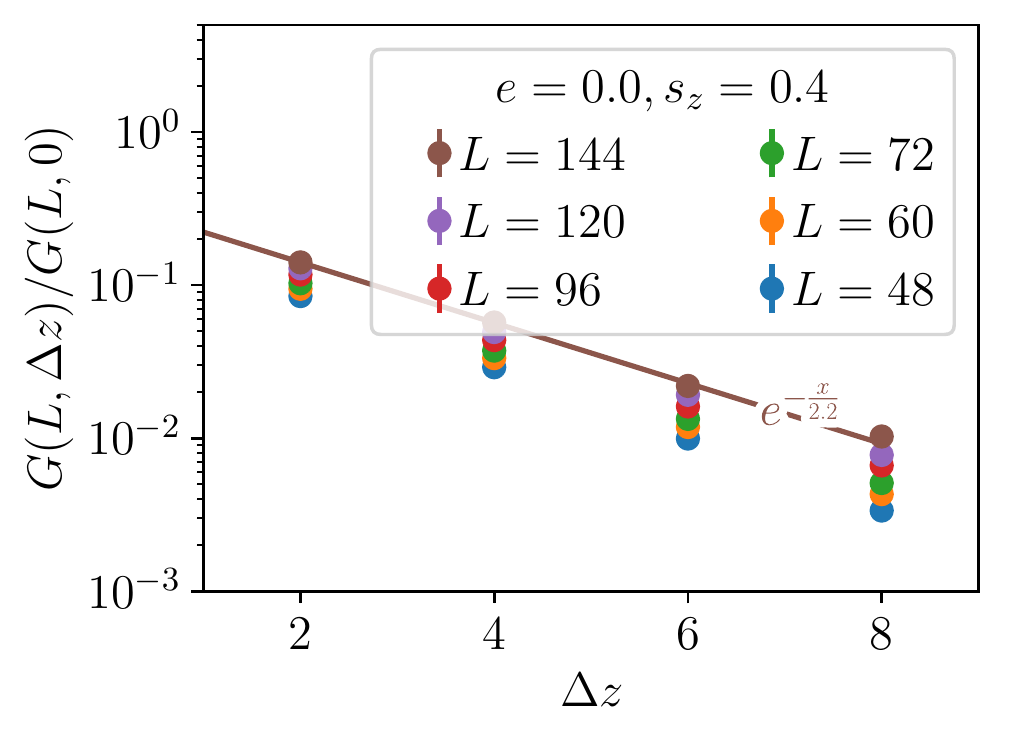}
\caption{The normalized inter-slab correlator of the integrated columnar order parameter (integrated over each slab) decays exponentially to zero with inter-slab distance $\Delta z$, but data at a fixed nonzero $\Delta z$ do not appear to converge to a limiting thermodynamic value even at the largest size $L$ accessible to our numerics.} 
\label{Fig:zdependenceofG}
\end{figure}

\begin{figure}
  \begin{subfigure}[b]{\columnwidth}
    \includegraphics[width=\linewidth]{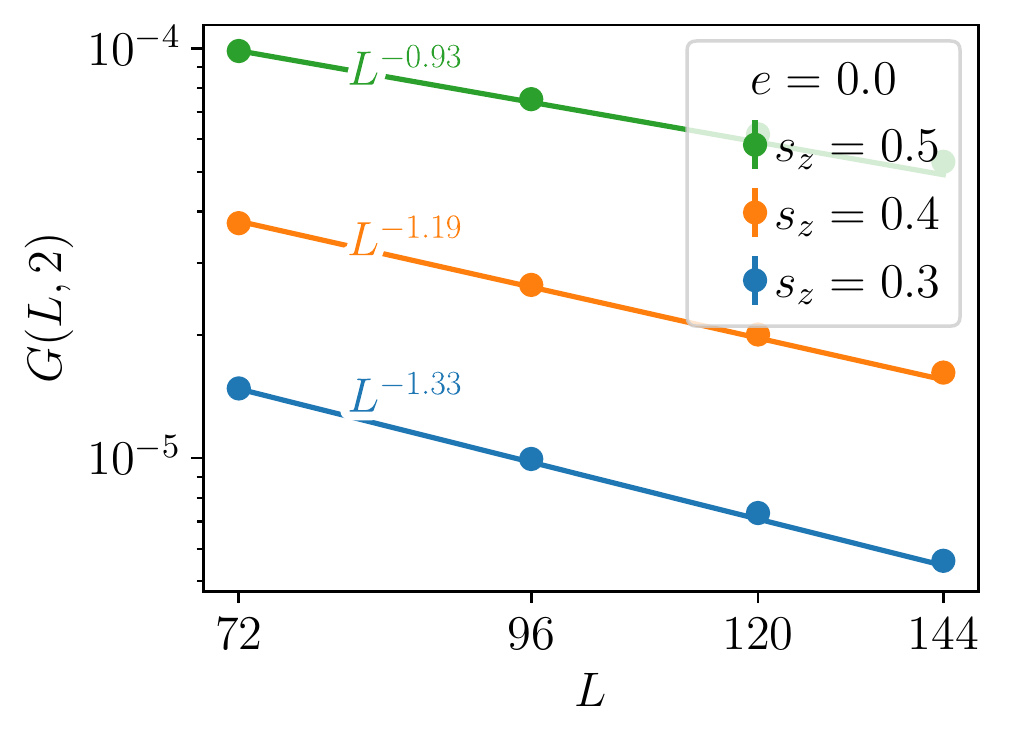}
  \end{subfigure}
  \hfill 
  \begin{subfigure}[b]{\columnwidth}
    \includegraphics[width=\linewidth]{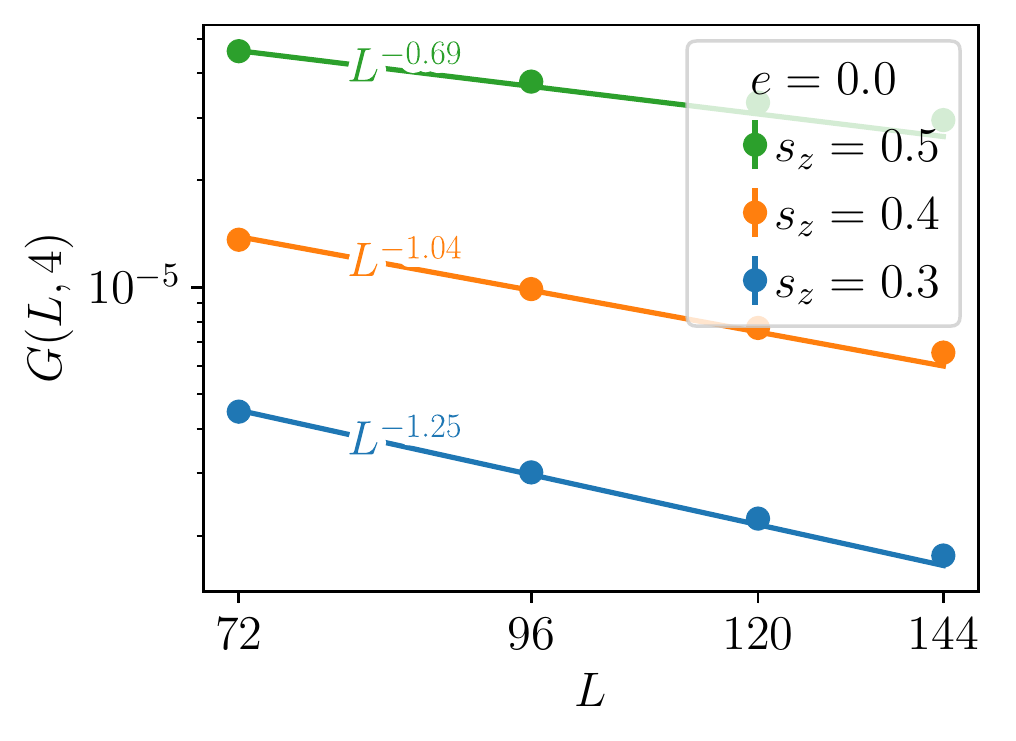}
  \end{subfigure}
  \caption{$G(L,2)$ (top) and $G(L,4)$ (bottom) as a function of $L$ on the $e=0$ cut for $s_z \in [0.3,0.5]$ is seen to decay as a power-law in $L$. This is the origin of the $L$-dependent drift seen in Fig.~\ref{zdependenceofG}}
     \label{Fig:Ginterlayer}
\end{figure}

Along the $e=0$ cut, we have $s_x=s_y=(3-s_z)/2$. Fig.~\ref{Fig:LCperp_e0LS_op} shows the $L$ dependence of the transverse columnar order parameter $\langle {\mathcal C}_x^2 +{\mathcal C}_y^2 \rangle$ and the transverse layering order parameter $\langle {\mathcal L}_x^2 +{\mathcal L}_y^2 \rangle$ for various $s_z$ in the vicinity of the transition from the $z$ layered phase to the sublattice ordered phase.
Both of these can both be used to distinguish between these two phases, with both tending to zero in the thermodynamic limit in the $z$ layered phase, and both tending to nonzero thermodynamic limits in the sublattice ordered phase. 

These order parameters also identify an interesting aspect of the $z$ layered phase, namely the presence of power-law order in the transverse columnar order parameter and the transverse layering order parameter. This is readily seen from the power-law fall off $\sim 1/L^{\eta_{\rm 3d}}$ of these quantities at large $L$, with $1< \eta_{\rm 3d} < 2$ and varying smoothly with $s_z$ within the layered phase. To explore this further, we have quantified the strength of two-dimensional columnar order within each occupied slab by measuring $G(L)$ as well as the connected correlation function $g(r)$ evaluated at $L/a$, where $a=4$. As is clear from Fig.~\ref{Fig:G(L)andg(r)}, within each slab, there are critical correlations corresponding to two-dimensional power-law columnar order within each occupied slab, with exponent $\eta_{\rm 2d}(0)$ that varies continuously with $s_z$.

If the two-dimensional columnar order parameters of different occupied slabs had only been coupled to each other via weak correlations decaying exponentially with increasing $\Delta z$, the measured $\eta_{\rm 3d}$ would have satisfied the equality  $\eta_{\rm 3d} = 1+\eta_{\rm 2d}(0)$. However, we find that the measured $\eta_{\rm 3d}$ and $\eta_{\rm 2d}(0)$ always satisfy $\eta_{\rm 3d} < 1+\eta_{\rm 2d}(0)$. To understand this better, we have also measured
$G(L,\Delta z)$ for various $\Delta z$ and $L$ in the layered phase. As is clear from Fig.~\ref{Fig:zdependenceofG}, the normalized inter-slab correlator $G(L,\Delta z)/G(L,0)$ does indeed appear to fall off exponentially with $\Delta z$ at fixed $L$. However, at fixed nonzero $\Delta z$, data for different $L$ do not appear to converge to a thermodynamic limit even at the largest $L$ accessible to our numerical study. This is seen most clearly in Fig.~\ref{Fig:Ginterlayer},  which shows the $L$ dependence of $G(L,\Delta z)$ for $\Delta z = 2, 4$ at various $s_z$ in the layered phase. As is clear from this figure, this quantity decays as a power law with increasing $L$ for fixed $\Delta z$. Moreover, the corresponding exponent $\eta_{\rm 2d}(\Delta z)$ {\em depends on $\Delta z$} in addition to its dependence on the fugacity $s_z$.
These transverse power-law correlations point to the unusual and interesting nature of this layered phase, which is discussed in more detail in Sec.~\ref{Sec:CriticalityandEFT}.

\begin{figure}
  \begin{subfigure}[b]{\columnwidth}
    \includegraphics[width=\linewidth]{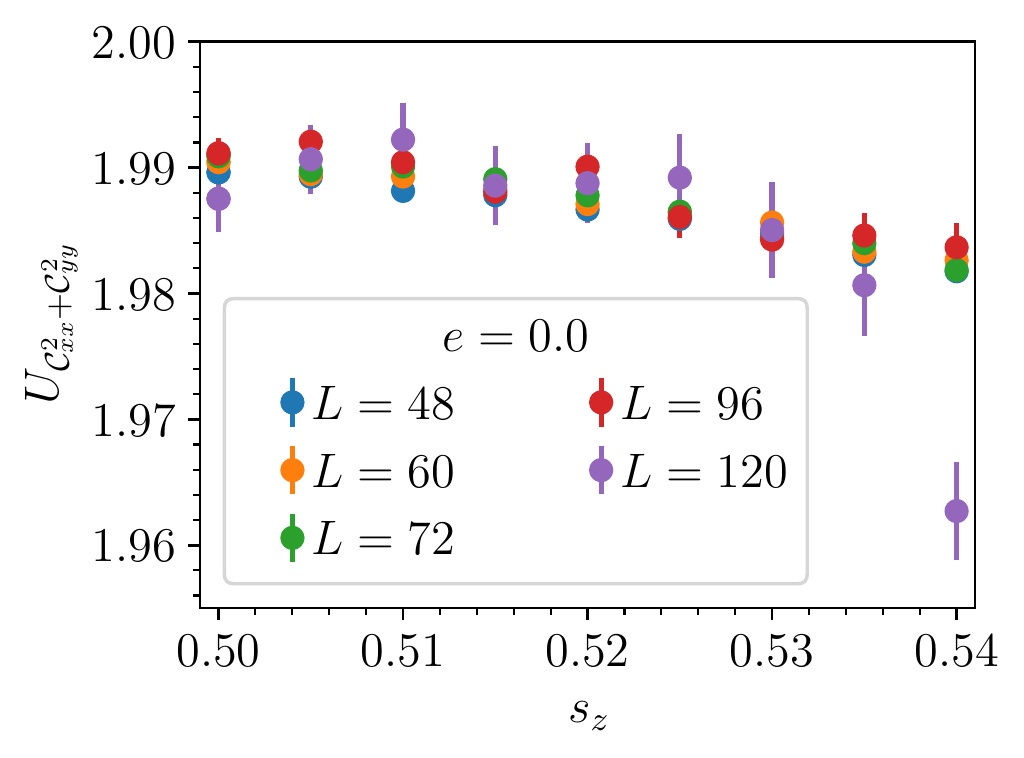}
  \end{subfigure}
  \hfill 
  \begin{subfigure}[b]{\columnwidth}
    \includegraphics[width=\linewidth]{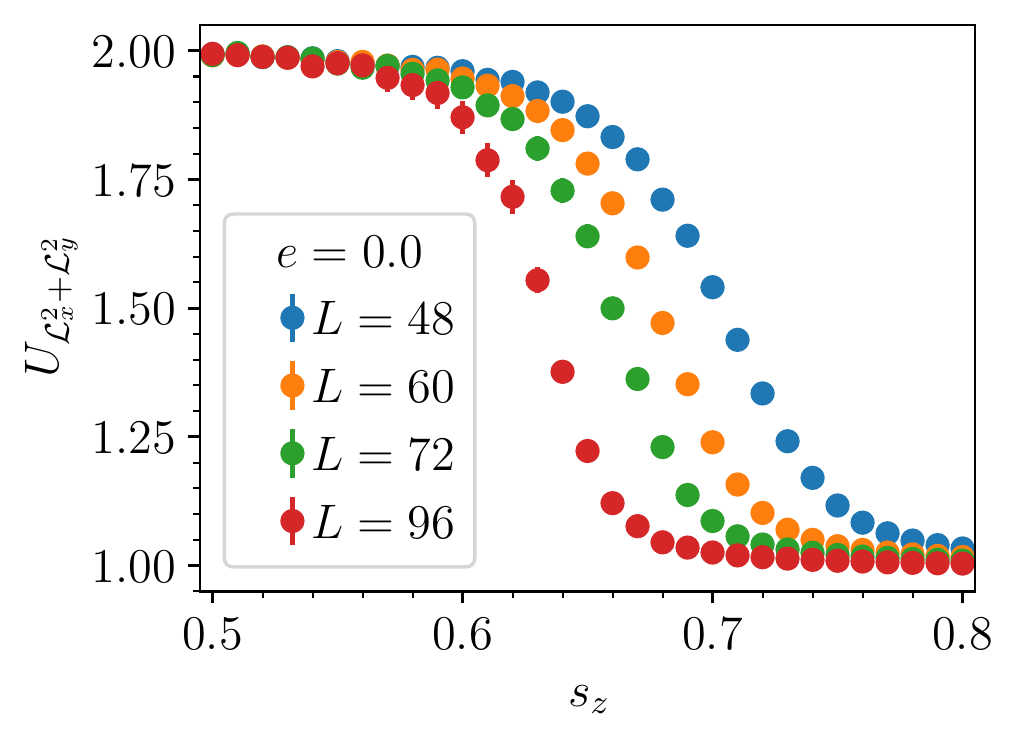}
  \end{subfigure}
  \caption{$U_{C_{xx}^2+C_{xx}^2}$ (top) and $U_{L_{x}^2+L_{y}^2}$ (bottom) as a function of $s_z$ on the $e=0$ cut across the $z$ layered to sublattice order transition for various $L$.}
     \label{Fig:LCperp_e0LS_bind}
\end{figure}

\begin{figure}[t]
	\includegraphics[width=\columnwidth]{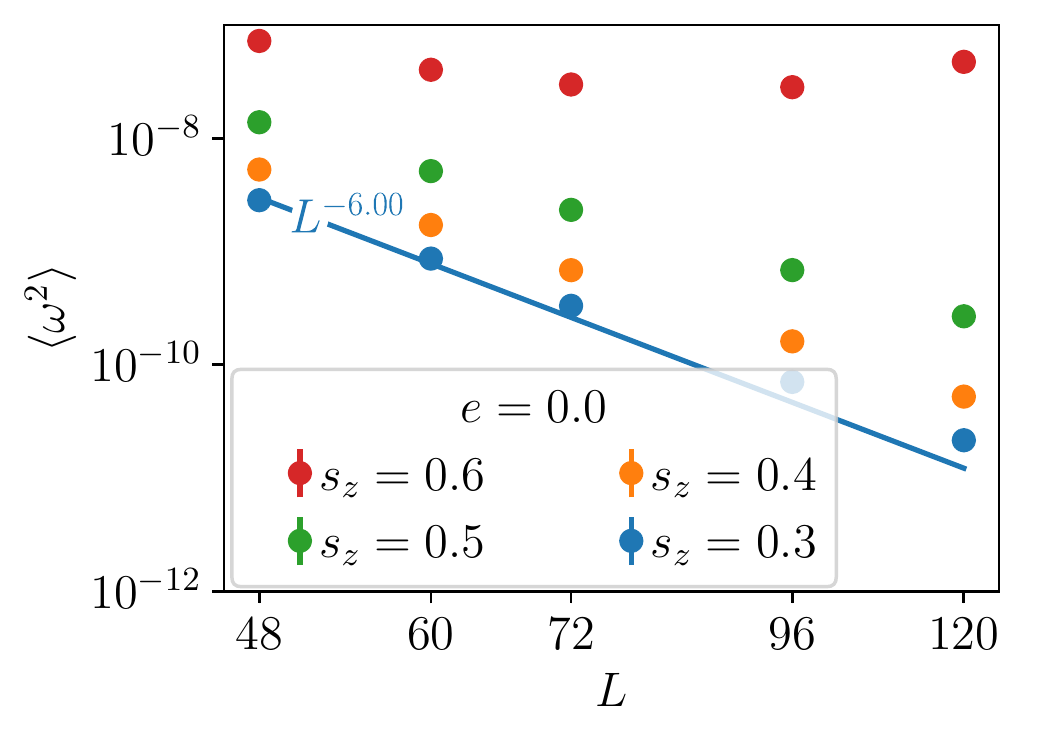}
	\caption{\label{Fig:omega_e0LS_bind} $\langle \omega^2 \rangle$ as a function of $L$ on the $e=0$ cut across the $z$ layered to sublattice order transition for $s_z \in [0.3,0.6]$. Note that $\langle \omega^2 \rangle$ falls off with $L$ somewhat slower than $1/L^6$ due to the presence of transverse power-law order within the occupied slabs.
	}
\end{figure}

To obtain a more precise estimate of the location of the transition between this $z$ layered phase and the sublattice ordered phase along the $e=0$ cut, we monitor the $s_z$ dependence of corresponding Binder ratios of these transverse order parameters for various sizes $L$. A conventional second order critical point between the $z$ layered phase and the sublattice ordered phase is expected to yield a crossing of the Binder ratio curves corresponding to different sizes $L$, whereas a first order transition is expected to lead to non-monotonic behaviour of these curves near the transition. As is clear from Fig.~\ref{Fig:LCperp_e0LS_bind}, we see neither of these behaviors. Instead, the Binder ratios for different sizes seem to stick together for $s_z \lesssim 0.535$, and splay apart for $s_z \gtrsim 0.535$. 

This sticking of the Binder ratio curves corresponding to different $L$ is what one expects in a critical phase, consistent with our earlier observation of power-law decay of the transverse layering and columnar order parameters as a function of system size $L$ in the $z$ layered phase. Further, the behavior of the Binder ratios in Fig.~\ref{Fig:LCperp_e0LS_bind} is reminiscent of the stick and splay behavior of the Binder of the columnar order parameter in the vicinity of the transition between long-range columnar order and power-law columnar order in the two-dimensional fully-packed lattice gas of hard squares and dimers. All of this points to the fact that the $z$ layered phase studied along this cut is indeed a critical phase. 

Finally, we note that the behavior of the layering and columnar order parameters in the vicinity of the isotropic point does not by itself provide unambiguous evidence of sublattice ordering (as opposed to columnar ordering). To obtain a direct confirmation of the sublattice-ordered nature of the phase, we have therefore studied the $L$ dependence of the $\langle \omega^2 \rangle$, which goes to zero with increasing $L$ in the $z$ layered phase but is seen to remain nonzero in the thermodynamic limit in the vicinity of the isotropic point, as seen in Fig.~\ref{Fig:omega_e0LS_bind}.

\begin{figure}
  \begin{subfigure}[b]{\columnwidth}
    \includegraphics[width=\linewidth]{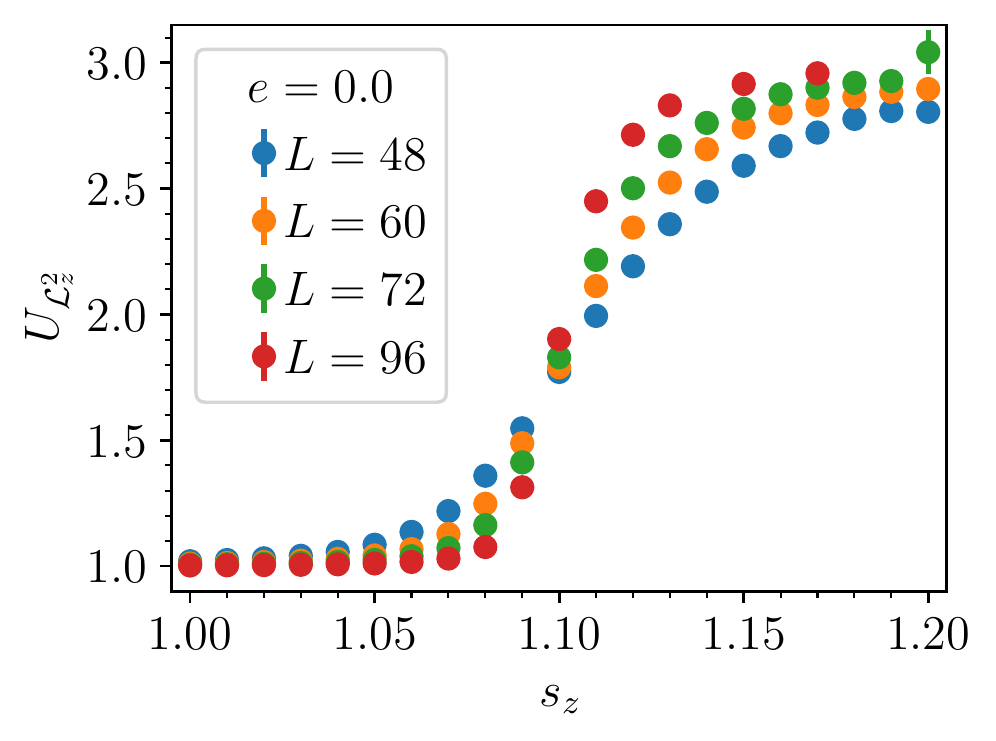}
  \end{subfigure}
  \hfill 
  \begin{subfigure}[b]{\columnwidth}
    \includegraphics[width=\linewidth]{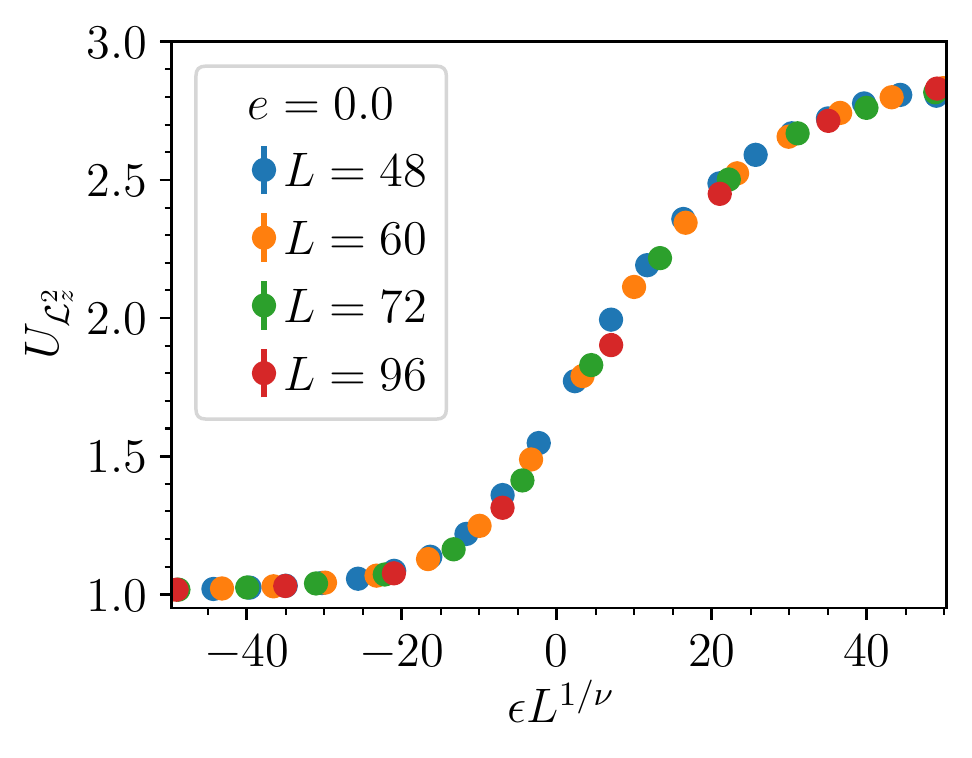}
  \end{subfigure}
  \caption{(top) $U_{L_{z}^2}$  as a function of $s_z$ on $e=0$ cut across the sublattice order to $z$ columnar transition for various $L$. (bottom) Scaling collapse of $U_{L_{z}^2}$ as a function of $\epsilon L^{1/\nu}$ with $\epsilon=(s_z-1.095)$ and $\nu \approx 0.63$ of the 3D Ising universality class~\cite{3dIsingexponent}.}
     \label{Fig:Lz_e0SC_bind}
\end{figure}

\begin{figure}
  \begin{subfigure}[b]{\columnwidth}
    \includegraphics[width=\linewidth]{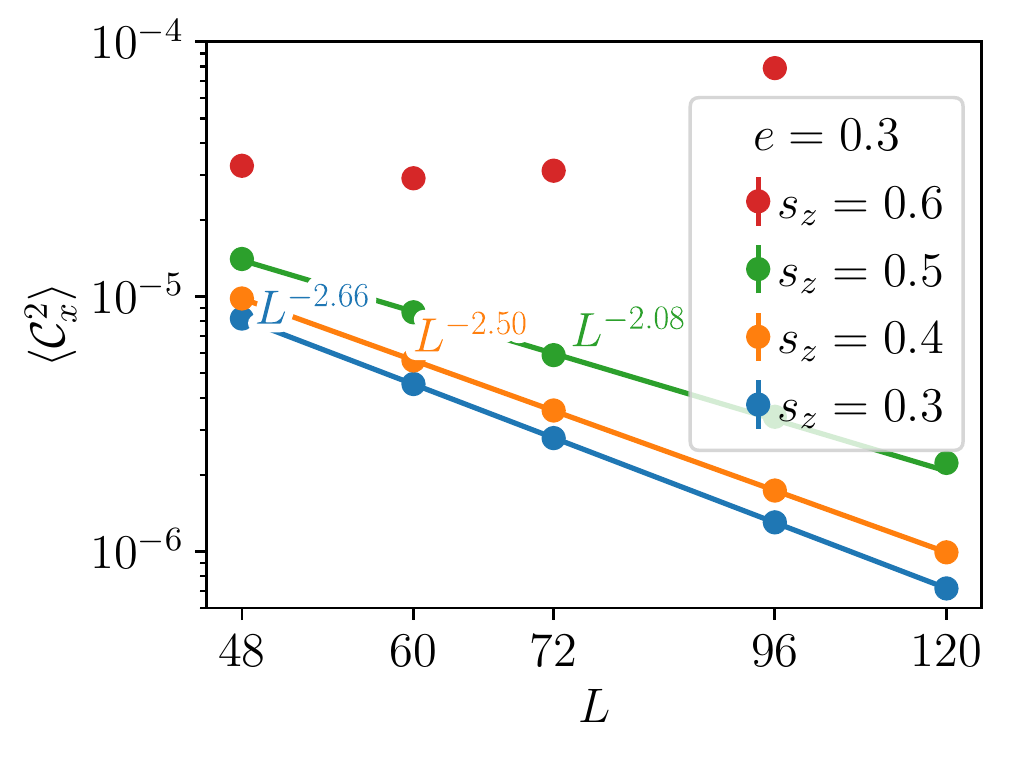}
  \end{subfigure}
  \hfill 
  \begin{subfigure}[b]{\columnwidth}
    \includegraphics[width=\linewidth]{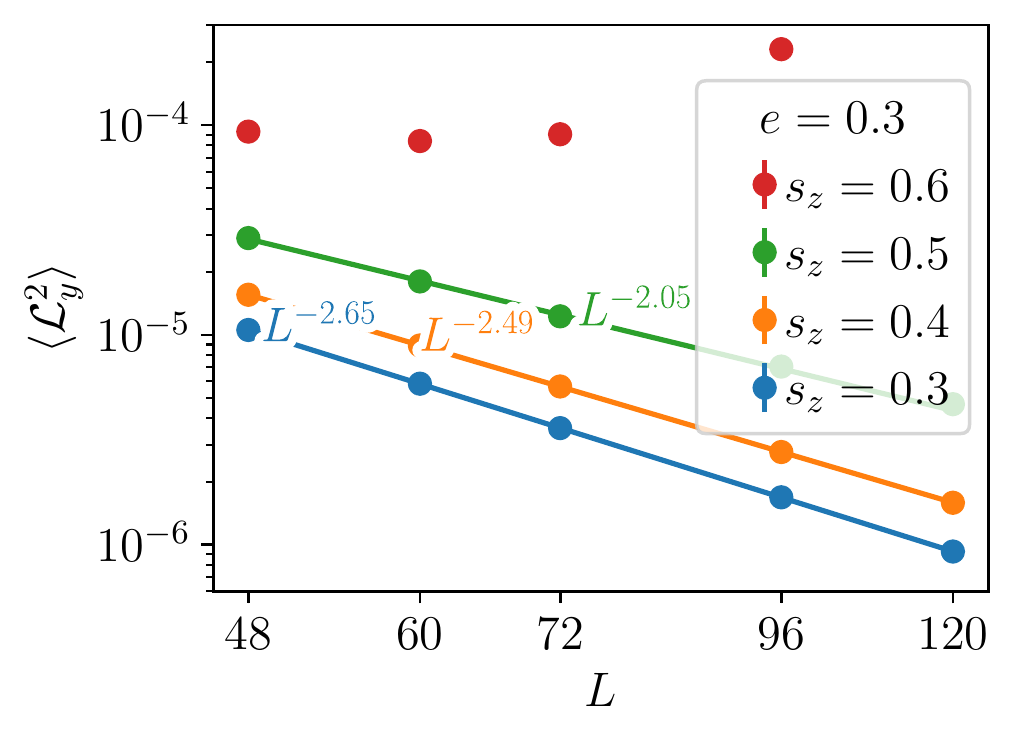}
  \end{subfigure}
  \caption{$\langle C_{x}^2 \rangle$ (top) and $\langle L_{y}^2 \rangle$ (bottom) as a function of $L$ on the $e=0.3$ cut across the $z$ layered to $x$ columnar transition for $s_z \in [0.3,0.6]$.}
     \label{Fig:LyCx_e03LC_op}
\end{figure}

\begin{figure}
  \begin{subfigure}[b]{\columnwidth}
    \includegraphics[width=\linewidth]{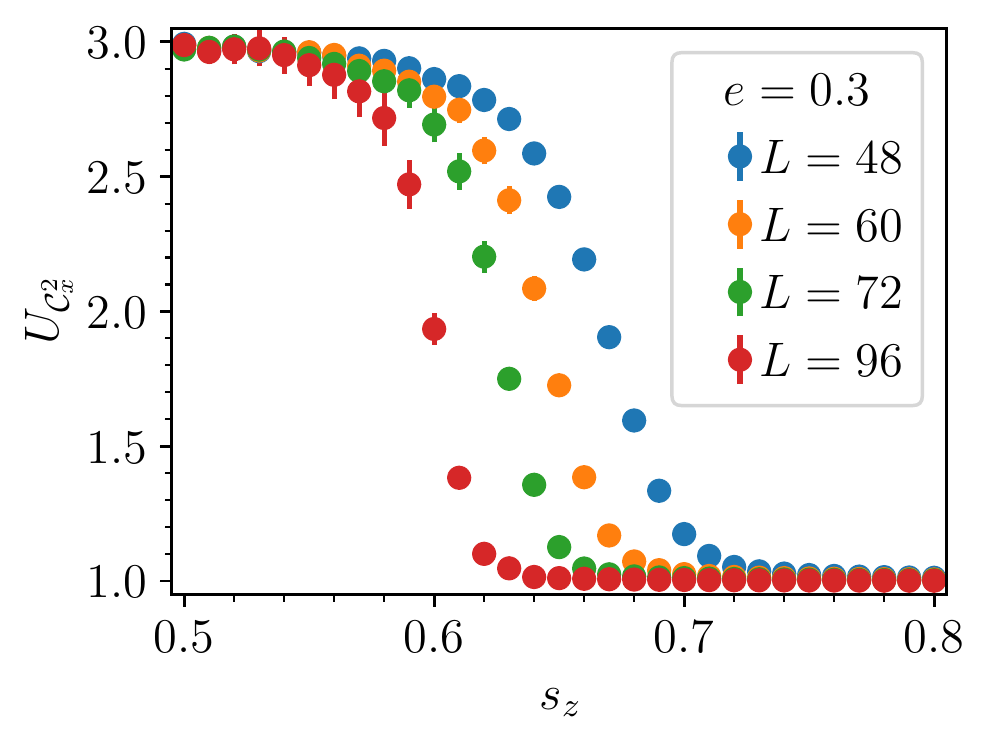}
  \end{subfigure}
  \hfill 
  \begin{subfigure}[b]{\columnwidth}
    \includegraphics[width=\linewidth]{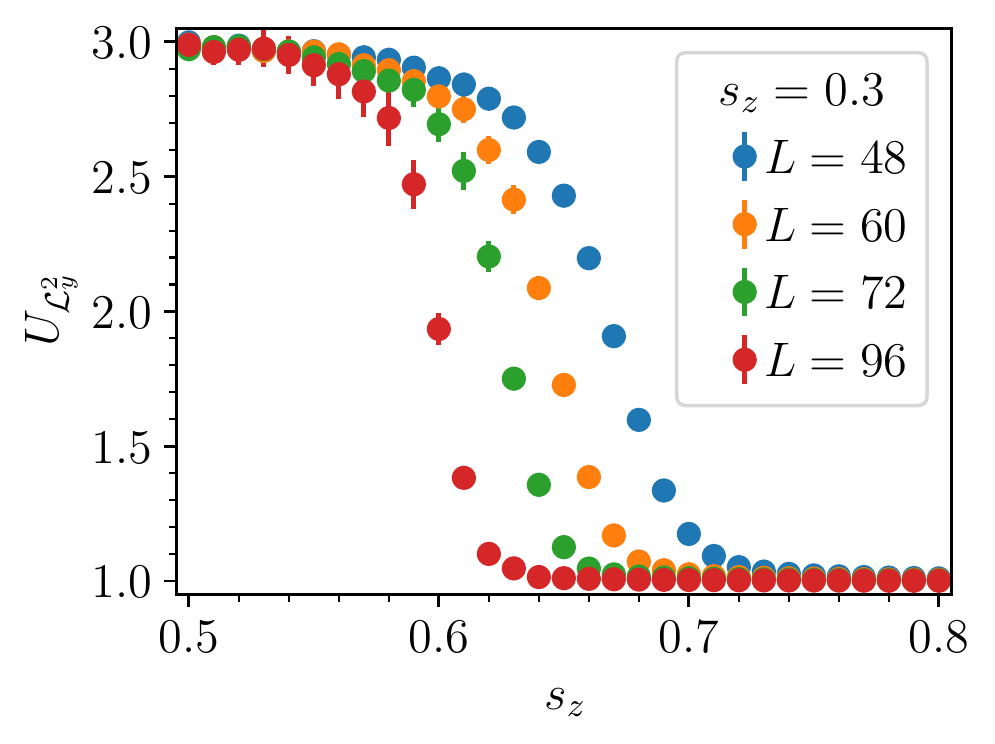}
  \end{subfigure}
  \caption{$U_{C_{x}^2}$ (top) and $U_{L_{y}^2}$ (bottom) as a function of $s_z$ on the $e=0.3$ cut across the $z$ layered to $x$ columnar transition for various $L$.}
     \label{Fig:LyCx_e03LC_bind}
\end{figure}

\begin{figure}
  \begin{subfigure}[b]{\columnwidth}
    \includegraphics[width=\linewidth]{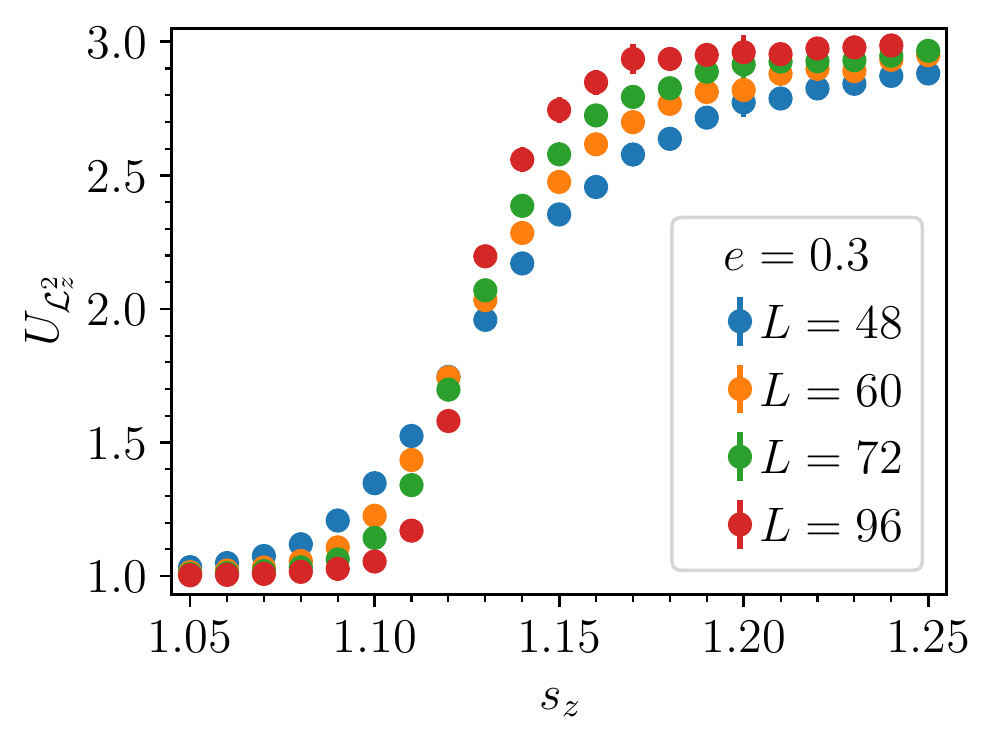}
  \end{subfigure}
  \hfill 
  \begin{subfigure}[b]{\columnwidth}
    \includegraphics[width=\linewidth]{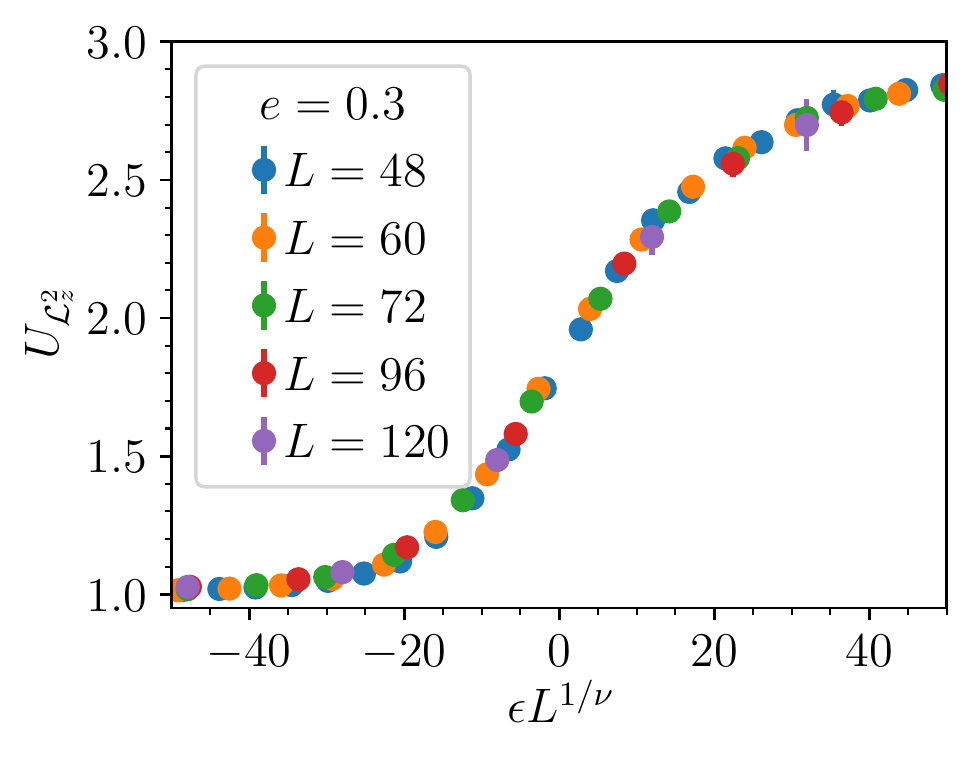}
  \end{subfigure}
  \caption{(top) $U_{L_{z}^2}$  as a function of $s_z$ on $e=0.3$ cut across the sublattice order to $z$ columnar transition for various $L$. (bottom) Scaling collapse of $U_{L_{z}^2}$ as a function of $\epsilon L^{1/\nu}$ with $\epsilon=(s_z-1.124)$ and $\nu \approx 0.63$ of the 3D Ising universality class~\cite{3dIsingexponent}.}
     \label{Fig:Lz_e03SC_bind}
\end{figure}

As $s_z$ is increased further along the $e=0$ cut, the sublattice-ordered phase terminates in a transition to $z$ columnar order, which breaks lattice translation symmetry in the $x$ and $y$ directions, but has full lattice translation symmetry in the $z$ direction. This is clear from examining the behaviour of $\langle {\mathcal L}_z^2 \rangle$ and the corresponding Binder cumulant. Since the $Z_2$ symmetry of lattice translations in the $z$ direction is restored upon crossing this transition, while translation symmetry in other directions remains broken on either side of the transition, one expects this transition to be in the three dimensional Ising universality class. Indeed, as is clear from Fig.~\ref{Fig:Lz_e0SC_bind} results for the Binder ratio and its scaling collapse in the vicinity of this transition, our results are entirely consistent with this expectation.

Along the $e=0.3$ cut, we have $s_x=s_y+0.3$, $s_y = (2.7-s_z)/2$. As we increase $s_z$ along this cut, we go from the $z$ layered phase first to an $x$ columnar phase, which has two-fold symmetry breaking of lattice translations in the $y$ and $z$ directions, while preserving lattice translation symmetry in the $x$ direction. This is apparent from the $L$ dependence of $\langle {\mathcal C}_x^2 \rangle$ and $\langle {\mathcal L}_y^2 \rangle$, as shown in Fig.~\ref{Fig:LyCx_e03LC_op}. A more precise estimate of the critical value of $s_z$ is obtained from the stick and splay behavior of the corresponding Binder ratios, as shown in Fig.~\ref{Fig:LyCx_e03LC_bind}.

As $s_z$ is increased further along this $e=0.3$ cut, there is a transition from the $x$ columnar phase to the sublattice ordered phase. Again, one expects this transition to be in the Ising universality class since the two phases differ only by the $Z_2$ symmetry breaking of lattice translations in the $x$ direction. Our results are indeed consistent with this expectation. The same expectation is also borne out by our results for the next transition along this cut, at which the sublattice ordered phase gives way to the $z$ columnar phase. This is clear from Fig.~\ref{Fig:Lz_e03SC_bind} which shows the corresponding crossing of the Binder ratios and their scaling collapse. For even larger values of $s_z$ there is yet another transition, from the $z$ columnar phase to a $y$ layered phase. We do not display our results for this transition since they are entirely analogous to previously displayed results for the transition between the $z$ layered phase and the $x$ columnar phase.

Unlike for the Ising transitions between the sublattice ordered phase and various columnar phases, we do not have a detailed understanding of the universality class of the transitions between the various layered phases and the sublattice ordered phase, and between the various layered phases and columnar phases. This is because of the unusual nature of the layered phase, which displays power-law transverse ordering perpendicular to the layering direction. What is the correct coarse-grained field-theoretical description of universal features of these continuous transitions? Although we are unable to answer this question, we note that the eventual answer would need to build on the answer to a related question: How can we understand the transverse power-law order of the layered phase itself within a coarse-grained effective field theory approach? Although our work does not fully answer this question either, the broad contours of an answer are sketched in the discussion below.

\section{Power-laws in the layered phase}
\label{Sec:CriticalityandEFT}
Each occupied slab in the $z$ layered phase is power-law ordered in the transverse directions, with power-law correlations of ${\mathcal L}_x$ and ${\mathcal L}_y$. To understand this better, we start in the $s_z \to 0$ limit.
When $s_z =0$, there are only $x$ and $y$ plates present, and these clearly organize themselves into occupied slabs  separated from each other by one lattice spacing along the $z$ axis. Viewed along the $z$ axis, each such occupied slab is seen to be equivalent to a fully-packed square lattice dimer model, with $x$ plates represented by dimers along $y$ links of the equivalent square lattice, and $y$ plates represented by dimers along $x$ links of this square lattice. The transverse power-law order of the occupied layers is simply understood in this limit as being a consequence of the power-law columnar order of the fully-packed dimer model on the square lattice. Thus, at $s_z=0$, we expect power-law columnar order with power-law exponent $\eta_{\rm 2d} = 2$.

As mentioned in  Sec.~\ref{Introduction}, the power-law correlations of the fully-packed square lattice dimer model can be understood via a coarse-grained field theory whose action is written in terms of a fluctuating scalar height field that represents the electrostatic potential of a system of fluctuating dipoles:
\begin{eqnarray}
S_{\rm 2d} &=& \pi g \int d^2 r (\nabla_{\perp} h)^2, \nonumber \\
Z &\propto & \int {\mathcal D}h \exp(-S),
\end{eqnarray}
with $g=1/2$.
In this description the local density of $x$ dimers (which represent $y$ plates) is given by $(-1)^{x+y}\partial_y h + {\mathcal A} (-1)^x\cos(2 \pi h)$, while the local density of $y$ dimers (which represent $x$ plates) is given by
 $(-1)^{x+y+1}\partial_x h + {\mathcal A} (-1)^y\sin(2 \pi h)$.

When $s_z = \epsilon$, with $\epsilon >0$ but very small, $z$ plates are allowed. However, and this is key, the only configurations that contribute to the partition function in the thermodynamic limit have {\em pairs} of $z$ plates, stacked one above the other in the $z$ direction. When two such $z$ plates are stacked one above the other in an occupied slab, they can be represented in the equivalent two dimensional system  by a small admixture of hard squares introduced into the fully-packed square lattice dimer model. The fugacity of these hard squares is of order $\epsilon^2$. This small density of hard squares increases the stiffness of the fluctuating height field, so that $g(\epsilon) = 1/2 + {\mathcal O}(\epsilon^2)$, leading to a decrease in the value of $\eta_{\rm 2d}$, so that $\eta_{\rm 2d} = 2 - {\mathcal O}(\epsilon^2)$~\cite{ramola_columnar_2015}.

If such a pair of $z$ plates straddles two neighbouring occupied slabs, it can be replaced by a {\em pair} of parallel $x$ or $y$ plates that straddle the two occupied slabs. To ${\mathcal O}(\epsilon^2)$, these are the leading defects that destroy perfect layering and couple neighbouring occupied slabs. Crucially, single $x$ or $y$ plates straddling neighbouring occupied slabs do not contribute in the thermodynamic limit. Such plates can be viewed as a pair of holes in one layer of each of the two adjacent occupied slabs. The full-packing constraint ensures that this pair of holes, which is a dipolar defect from the point of view of the equivalent problem of hard squares and dimers, cannot move on its own. In effect, the full-packing constraint ensures that these dipolar defects come in nearest-neighbour pairs. Thus, dipolar defects are confined into quadrupoles. This is very different from the purely two-dimensional problem of hard-squares and dimers. The fact that each occupied layer is actually part of a fully three-dimensional system is therefore crucial for understanding the nature of the defects, although our description is in terms of an equivalent two dimensional system.

We now argue that this confinement of the dipolar defects into quadrupoles is the reason that the power-law correlations within each occupied slab survive the coupling between neighbouring occupied slabs. To see this, represent each occupied slab in the small $\epsilon$ limit as a two-dimensional system of hard-squares and dimers.
Ignoring the coupling between neighbouring occupied slabs for the moment, this can be described by the effective action:
\begin{eqnarray}
S_0 &=& \pi g \sum_z \int d^2 r (\nabla_{\perp} h_z)^2.
\end{eqnarray}
 The quadrupolar defects that couple the neighbouring occupied slabs can then be represented by attractive interactions between one pair of parallel dimers in a layer and another such pair in the neighbouring layer. This leads to terms in the effective action that couple the height fields of neighbouring slabs:
\begin{eqnarray}
S&=&S_0+S_1+S_2, \nonumber \\
S_1 &=& \lambda \sum_z \int d^2 r\left[ (\partial_x^2 h_z)(\partial_x^2 h_{z+1}) + (\partial_y^2 h_z)(\partial_y^2 h_{z+1}) \right]+ \nonumber \\
&&+ \lambda^{'} \sum_z \int d^2 r(\partial_{xy}^2 h_z)(\partial_{xy}^2 h_{z+1}) , \nonumber \\
S_2 &=& \lambda^{''} \sum_z \int d^2 r \cos(4 \pi (h_z(r) -h_{z+1}(r))),
\end{eqnarray}
with $\lambda$, $\lambda^{'}$, and $\lambda^{''}$ all vanishing in the $\epsilon \to 0$ limit.
Along the renormalization group (RG) fixed line with action $S_0$ parameterized by $g$, all three perturbations that constitute $S_1$ and $S_2$ are seen to be {\em irrelevant} so long as $8/g > 4$, {\em i.e.} $g < 2$.
The quadrupolar nature of the defects coupling neighbouring power-law correlated slabs is now seen to be the key reason for the stability of these power-law correlated slabs at nonzero $s_z$: In the absence of the constraints that force the dipolar defects to be confined into quadrupoles, a cosine term of the form $\cos(2 \pi (h_z(r) -h_z(r+1)))$, representing a {\em single} $x$ or $y$ plate straddling two neighbouring occupied slabs, would have been allowed. This term is relevant unless $2/g> 4$, {\em i.e.} $g < 1/2$, which is never the case since $g=1/2$ for $s_z=0$, and a small ${\mathcal O}(\epsilon)$ value of $s_z$ only {\em increases} the value $g$ from $g=1/2$ to $g = 1/2+ {\mathcal O}(\epsilon^2)$.
Thus, the transverse power-law correlations in this layered phase  are stabilized by constraints imposed by full-packing, which confine dipolar inter-slab defects into quadrupoles. 
\begin{figure}
\includegraphics[width=\linewidth]{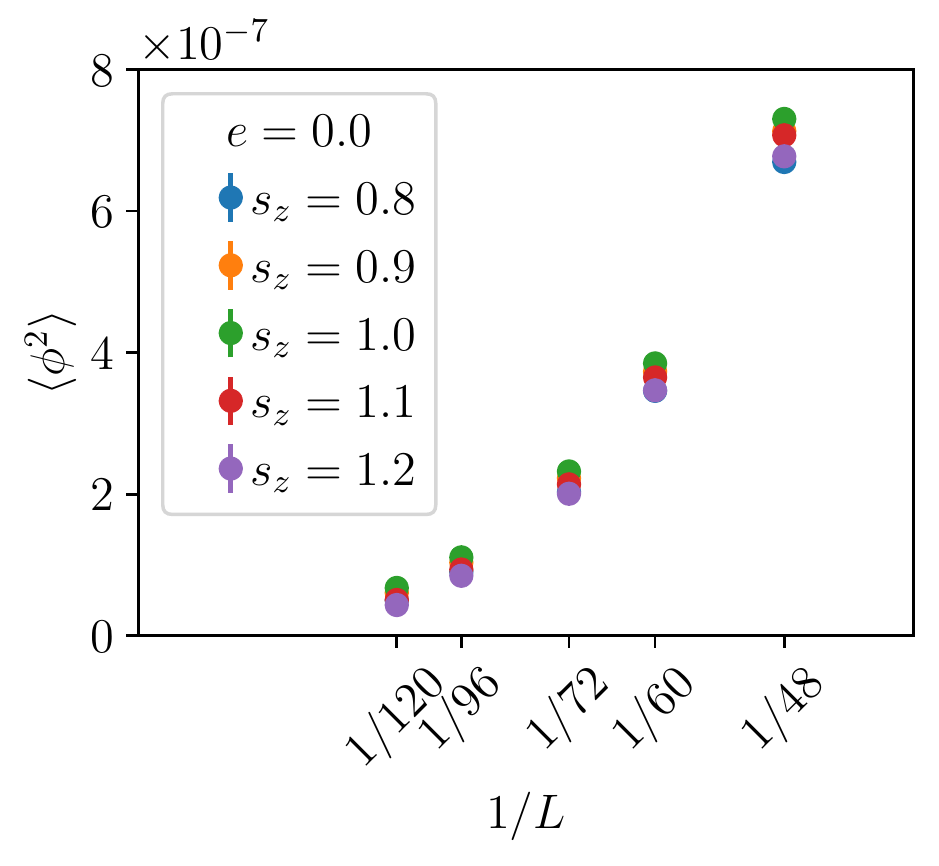}
\caption{The mean-square scalar sublattice order parameter $\langle \phi^2 \rangle$ goes to zero in the thermodynamic limit even in the sublattice-ordered phase.} 
\label{phisquaregoestozero}
\end{figure}

\section{Outlook}
\label{sec:Outlook}
We conclude by highlighting some unresolved questions that may be worth addressing in follow-up studies.
The first question has to do with our choice of sublattice order parameter Eq.~\ref{usefulsublatticedefn}.
One might a priori expect that the following alternate definition of a scalar sublattice order parameter provides a more natural and straightforward probe of sublattice order:
\begin{equation}
\phi = \frac{1}{L^3}\sum_i (-1)^{x(i)+y(i)+z(i)}(\eta_{xx}(i)+\eta_{yy}(i)+\eta_{zz}(i)) \; .
\end{equation}
Our results indicate that $\langle \phi^2 \rangle$ defined in this way goes to zero in the thermodynamic limit even in the sublattice-ordered phase, and cannot be used to draw any reliable conclusions. This is illustrated in Fig.~\ref{phisquaregoestozero}. Why this is the case is not entirely clear. Since $\omega$ can in any event be used to unambiguously detect the presence of sublattice order independent of this curious behaviour of $\phi$, we may safely side-step this difficulty as far as our results here are concerned. However, we emphasize that this behaviour of $\phi$ is an interesting puzzle that deserves further study in follow-up work.

The second set of questions has to do with the coarse-grained description discussed above, which  provides a natural explanation of the power-law correlations within each occupied slabs in the layered phase. It should be possible to develop this analysis further, in order to also describe the nature of the correlations between two different occupied slabs separated by a distance $\Delta z$ in the layering direction. In particular, it would be interesting to develop a theory for  the relationship between the intra-slab power-law correlation exponent and the power-law exponent characterizing the $L$ dependence of the three-dimensional transverse layering order parameter $\langle {\mathcal L}_x^2 + {\mathcal L}_y^2 \rangle$. Also worth pursuing is the effect of a small repulsive interaction between $x$ and $y$ plates in this layered phase with $s_z \ll s_x= s_y$. If this preserves the layered phase, it would be interesting to explore its effects on the transverse columnar ordering. 

Finally, we note that we have also explored in closely-related work~\cite{Dipanjan2021Vacancy} a somewhat different layered phase, accessed by introducing vacancies into the isotropic fully-packed system. It would be interesting to ask how this layered phase is related to the one discussed here, perhaps by studying the effect of a small density of vacancies in the limit of $s_z \ll s_x= s_y$.

\section{Acknowledgements}
We acknowledge useful discussions with R. Dandekar, R. Moessner, K. Ramola, and N. Shannon. GR gratefully acknowledges technical assistance from K. Ghadially and A. Salve of the Department of Theoretical Physics (DTP) of the Tata Institute of Fundamental Research (TIFR) and from the Scientific Computing and Data Analysis section of the Okinawa Institute of Science and Technology (OIST). A major portion of the results presented here formed part of the Ph.D thesis submission of GR to the TIFR Deemed University (2019), and was made possible by the generous allocation of computational resources made by DTP TIFR. The final stages of this work were facilitated by extensive computational resources provided by OIST. GR was supported by a graduate fellowship of the TIFR during a major part of this work, and by the TQM unit at the Okinawa Institute of Science and Technology during the final stages of this work. The work of SB was supported by a postdoctoral fellowship of the DTP TIFR. DM was supported by a graduate fellowship of the Institute of Mathematical Sciences (IMSc). KD was supported at the
TIFR by DAE, India and in part by a J.C. Bose Fellowship
(JCB/2020/000047) of SERB, DST India, and by
the Infosys-Chandrasekharan Random Geometry Center
(TIFR). DD’s work was partially supported by Grant No. DST-SR-
S2/JCB-24/2005 of the Government of India, and partially by a Senior Scientist Fellowship of the National Academy of Sciences.

{\em Author contributions:} GR performed the computational work with assistance from DM. KD, RR, and DD conceived the project, directed the computational work, and finalized the manuscript using detailed inputs from GR. SB performed exploratory simulations in the preliminary stages of the work.

\bibliography{references}

\end{document}